\documentclass[referee,a4paper,fleqn,usenatbib]{mnras}

\usepackage[T1]{fontenc}
\usepackage{ae,aecompl}
\usepackage[normalem]{ulem}

\usepackage{graphicx}	

\newcommand{\pone}{{PKS\,0213$-$13}}
\newcommand{\ptwo}{{PKS\,0230$-$10}}
 
\newcommand{\pfour}{{NVSS\,J223932-042933}} 
\newcommand{\pfive}{{PKS\,0215$-$18}} 

\newcommand{\appropto}{\mathrel{\vcenter{
  \offinterlineskip\halign{\hfil$##$\cr
    \propto\cr\noalign{\kern2pt}\sim\cr\noalign{\kern-2pt}}}}}

\title[LEAP: Direction Dependent Ionospheric Calibration for Low Frequencies]{LEAP: An Innovative Direction Dependent Ionospheric Calibration Scheme for Low Frequency Arrays}

\date{Accepted XXX. Received YYY; in original form ZZZ}

 \author[M. Rioja et al.]{
  Mar\'{\i}a J. {Rioja},$^{1,2,3}$\thanks{Email: maria.rioja@icrar.org}
  Richard {Dodson}$^1$,
  Thomas M.O. {Franzen}$^{4,2}$
\\
 {$^1$ ICRAR, M468, The University of Western Australia, 35 Stirling Hwy, Crawley, Western Australia, 6009}\\
 {$^2$ CSIRO Astronomy and Space Science, PO Box 1130, Bentley WA 6102, Australia}\\
 {$^3$ Observatorio Astron\'omico Nacional (IGN), Alfonso XII, 3 y 5, 28014 Madrid, Spain}\\ 
 {$^4$ International Centre for Radio Astronomy Research, Curtin University, Bentley, WA 6102, Australia}\\
}
\date{Accepted XXX. Received YYY; in original form ZZZ}

\begin{document}
\label{firstpage}
\pagerange{\pageref{firstpage}--\pageref{lastpage}}
\maketitle

\begin{abstract} 

The ambitious scientific goals of the SKA require a matching capability for calibration of atmospheric propagation errors, which contaminate the observed signals. 
We demonstrate a scheme for correcting the direction-dependent ionospheric and instrumental phase effects at the low frequencies and with the wide fields of view planned for SKA-Low. 
It leverages bandwidth smearing, to filter-out signals from off-axis directions, allowing the measurement of the direction-dependent antenna-based gains in the visibility domain; by doing this towards multiple directions it is possible to calibrate  across wide fields of view. 
This strategy removes the need for a global sky model, therefore all directions are independent. 
We use MWA results at 88 and 154 MHz under various weather conditions to characterise the performance and applicability of the technique.
We conclude that this method is suitable to measure and correct for temporal fluctuations and  direction-dependent spatial ionospheric phase distortions on a wide range of scales: both larger and smaller than the array size. The latter are the most intractable and pose a major challenge for future instruments.
Moreover this scheme is an embarrassingly parallel process, as multiple directions can be processed independently and simultaneously. This is an important consideration for the SKA, where the current planned architecture is one of compute-islands with limited interconnects. 
Current implementation of the algorithm and on-going developments are discussed.
\end{abstract}










\section{Introduction}

The ionospheric propagation medium effects are the dominant source of errors in interferometric observations at low radio frequencies, and pose fundamental barriers to achieving the scientific objectives of the SKA.
They arise from the unpredictability in the spatial distribution of the total (free) electron content (TEC) in the 
ionosphere, 
and are responsible for changes in the path delays measured by an interferometer
in different sky directions (i.e. direction-dependent, {\it hereafter} DD). Moreover these can change on very short timescales of ca. few tens of seconds.
In general, GPS-based TEC maps do not have enough accuracy for ionospheric phase calibration due to lack of small spatial scale and short timescale TEC variation information.
Consequentially the corrections must be derived from the observations themselves, that is the data must be self-calibrated.

The ionospheric effects are increasingly dominant towards lower frequencies, where the magnitude of the ionospheric delay error is largest ($\tau_{ion} \propto \Delta {\rm TEC} \lambda^2$, where $\Delta {\rm TEC}$ is the difference in the {\rm TEC} observed from two antennas), and the instruments naturally have larger fields of view (FoVs), which limit the use of standard antenna-based direction-independent self-calibration techniques.
The residual ionospheric errors propagate, to a first order, into apparent local source position shifts in the images 
caused by TEC gradient phase structure over the array; the ionospheric errors are DD and the magnitude and direction of the shifts change across the FoV.
For a given source, those shifts can change with time, leading to coherence losses when combining multiple snapshot observations for increased sensitivity. 
Moreover 
higher order phase structure over the array will also cause defocusing or
deformation of the sources and signal loss in the individual snapshots.
Another challenge is in the calibration of the instrumental response, which requires precise knowledge of the complex antenna beam shapes.
Small differences between antenna beam models and the actual shapes result in errors (in general DD) in both phase and amplitude, which will affect the images.  

Calibration is a vital part of, but of course not the only challenge for, imaging in the SKA era. 
Wide-field imaging requires accounting for non-coplanar effects (i.e. {\it w}-terms) and applying DD calibration (i.e. time varying {\it A}-terms), which are 
the focus of intensive investigation \citep[e.g.][]{offringa2014,ddf_tasse}. 
Both of these challenges become more significant as a function of increasing wavelength, FoV and baseline lengths. 

The ambitious goals of the SKA (and its pathfinders) require the development of both advanced calibration and imaging techniques for low frequency arrays with large FoV; 
this is a very active field of research, driven forward by the SKA pathfinders.    
In this paper we  focus predominately on calibration issues.

The ``field based calibration'' \citep{cotton_04,kassim_07} was the first technique developed to correct for DD ionospheric phase errors, and was used for VLSS \citep{VLSS} and its follow up, VLSSr \citep{VLSSr}.
Field based calibration is limited to small size arrays relative to the scale of phase structure in the ionosphere, which in turn depends on the observing frequency and ionospheric conditions.
This is the case for all methods based in the image domain, since they can only correct for the first order ionospheric refractive shifts, measured from the image.
The technique known as ``peeling''  \citep{noordam_04,intema_09} addresses many of the calibration challenges; it provides solutions for a particular direction by subtracting the contributions from the rest of the sky. Therefore only the remaining source (or cluster of sources) contributes to the required solution. The calibration of the whole FoV is obtained using solutions for a discrete number of directions. It necessarily uses sequential processing 
so that the source in the direction being self-calibrated dominates over the errors from other directions and requires a global sky model  (i.e. a detailed model of all the sources in the FoV).
Other DD calibration schemes developed for low frequency observations build upon the ``peeling'' technique,
such as SPAM \citep{intema_09,intema_phd,intema_14} 
and ``Facet calibration'' \citep{vanweeren_16},
albeit with important differences among them \citep[see][and references therein for comparison]{vanweeren_16}.
Both Sagecal \citep{yatawatta_08,kazemi_11} and killMS \citep{killms, tasse_14b, tasse_14a} use an alternative approach, which is to solve for multiple directions simultaneously using an approximate matrix inversion. This can be a computationally efficient implementation,
however this approach does not use parallel processing for different directions and requires a large multi-node shared memory computer architecture. (We note that there is, however, some possibilities for parallelism over different frequency sub-bands.)

We present an end-to-end demonstration of a parallel DD calibration scheme in the visibility domain, {\it hereafter} Low frequency Excision of Atmosphere in Parallel (LEAP), 
that does not use ``peeling'', nor requires a global sky model, nor computationally expensive simultaneous solutions derived on large machines;
it builds upon previous discussions on frequency and time averaging in interferometric data \citep[e.g.][]{tms}. 
This is a routine consideration in wide-field imaging of VLBI data, where narrow frequency channels and small integration times are used to avoid limiting the FoV through decorrelation and smearing. 
Of interest here is the reverse application; to shrink the naturally wide FoVs of the next-generation low frequency arrays by averaging over the observed bandwidth, to  enable DD calibration.
This approach is embarrassingly parallel, because the response of the array has been highly beamed in the direction to be calibrated, and therefore multiple directions can be processed independently and simultaneously.

The development of the scheme, along with its characterisation, 
has been carried out using a subset of observations from the GaLactic and Extragalactic All-sky MWA survey \citep[GLEAM,][]{wayth_15}, conducted with Phase 1 of the Murchison Widefield Array \citep[MWA,][]{lonsdale_09,tingay_13}, including the lowest frequency band. The MWA is a low-frequency radio interferometer operating between 72 and 300 MHz, with an instantaneous bandwidth of 30.72 MHz. Phase 1 of the MWA consisted of 128 16-crossed-pair-dipole tiles ({\it hereafter} antennas), distributed over an area approximately 3 km in diameter. In this array configuration, using a  uniform weighting scheme, the angular resolution at 154 MHz was about 2.5 arcmin. The primary beam FWHM at this frequency is 27\degr, resulting in a huge FoV.
GLEAM covers the entire sky south of declination $+$30\degr\ and comprises observations at the following 5 frequency bands: 72--103, 103--134, 139--170, 170--200 and 200--231 MHz. 

The ionospheric calibration in the standard analysis of GLEAM \citep{nhw_gleam} 
was conducted by measuring apparent source position shifts in the 2-min snapshot images, compared to the catalogue positions in the 
NRAO VLA Sky Survey (NVSS, \citeauthor{condon_98} \citeyear{condon_98}) or  Molonglo Reference Catalogue (MRC, \citeauthor{large_81} \citeyear{large_81}).
A radial basis function was fitted to these shifts with a scale size of 10\degr, the typical scale size for patches with similar shifts. The radial basis function was then used to warp the original snapshot image.
Similar to the case of ``Field based calibration'', this image-based strategy is equivalent to correcting for the first-order linear distortions in the visibility domain  using the snapshot duration as the solution interval. 

The current LEAP implementation is the first step in our developments to address increasingly challenging situations 
arising from the longer baselines (i.e. in MWA Phase-2, LOw Frequency ARray (LOFAR) \citep{lofar} and  SKA-Low \citep{ska}).  
In  this  paper  we  describe  the  basis of the LEAP scheme in Section \ref{sec:leap}.
The MWA observations and analysis methods used to validate the scheme are presented in Section \ref{sec:obsmeth}. In Section \ref{sec:res}, we investigate the 
performance and feasibility of LEAP for DD calibration and to characterise the ionospheric distortions at fine scales. Finally we review our conclusions in Section \ref{sec:conc}.

\section{Basis of LEAP}\label{sec:leap}

Here we outline the LEAP strategy to measure and correct for DD phase errors ({\it hereafter} DDE)  in observations at low frequencies 
and with wide FoVs.
Both are carried out in the visibility domain.
The inputs to the DD data analysis are the direction independent ({\it hereafter} DI) calibrated visibility snapshot datasets (i.e. corrected for instrumental and global atmospheric effects). 

\subsection{Bandwidth smearing as a directional filter}

LEAP uses chromatic aberration to reduce the  FoV of the interferometer, and separate out incoming radiation from different directions. 
Averaging the visibilities over a finite bandwidth 
introduces aberrations in the image, which 
decreases the peak response and broadens the radial width of the sources.
This effect is called ``bandwidth smearing'' and is described further in \citet{bw_smearing}.
The degree of smearing is directly proportional to the angular distance from the 
phase-tracking centre. At a given angular distance, the effect is stronger for wider averaged bandwidths and, in an equivalent fashion, for longer baselines. The bandwidth smearing is independent of the observing frequency, for a given network of antennas.

Table \ref{table:sources} lists the corresponding attenuation for directions away from the phase centre, for MWA observations with a maximum baseline length equal to 3 km when averaging over the maximum instantaneous bandwidth of 30 MHz.
Since the phase-tracking centre can be changed to any position across the FoV using post-correlation phase rotation, in particular to a calibrator source position, this effectively acts as a directional filter to separate out the signal of only the calibrator.
The requirements to be a LEAP calibrator  are discussed in Section \ref{subsec:leapcals}.

While the efficiency of the  directional filtering will benefit from averaging across larger bandwidths, as shown in Table \ref{table:sources}, 
attention should be paid to ensure one does not lose significant information on the spectral signature of the errors. 
How much one can average without introducing significant errors depends on the baseline length, the observing frequency, and external factors like the weather conditions. 
This is discussed in detail in Section \ref{sec:error}.

\begin{table}
\centering
\begin{tabular}{cclr}
\hline
\hline
$\beta= {\Delta \nu \over \nu_o} {\theta_o \over \theta_{beam}}$ & $\theta_o$ & $\rho$ (\%loss) & $\Delta$w \\
\hline
0 & 0$^\prime$ & 1.00 (0\%) & 1.0 \\
1 & 10$^\prime$ & 0.81 (19\%) & 1.2 \\ 
2 & 20$^\prime$ & 0.52 (48\%) & 1.9\\ 
6  & 60$^\prime$ & 0.18 (82\%) & 5.6 \\ 
12  & 2\degr& 0.09 (91\%) & 11.2 \\ 
48  & 8\degr & 0.02 (98\%) & 45.1 \\ 
\hline
\hline
\end{tabular}
\begin{footnotesize}
\caption{``Bandwidth smearing'' image distortions for MWA observations 
with a bandwidth $\Delta \nu = 30$MHz, a fractional bandwidth $\Delta \nu / \nu_0 \sim 0.4$ and synthesised beamwidth $\theta_{beam} \sim 4'$ corresponding to a maximum baseline length of 3 km. 
{\it Column 1:} The dimensionless parameter, $\beta$, is used to characterise the bandwidth smearing effects. The attenuation is significant for source distances from the phase-tracking centre where $\beta > 1$. The value of $\beta$ is given by the product of the fractional bandwidth ($\Delta \nu \over \nu_o$)
and the source distance from the phase centre ($\theta_o$) in synthesised beamwidths ($\theta_{beam}$);
{\it Column 2:} distance of the object from the phase-tracking centre; {\it Column 3:} dimensionless parameter $\rho$, the peak response attenuation factor along with percentage brightness reduction in parenthesis; and {\it Column 4:} dimensionless parameter $\Delta$w, the factor by which the source width is increased along the radial direction. Note that the bandwidth smearing effect is stronger for longer baselines, but for a given baseline length or array it is independent of the observing frequency. \label{table:sources}}
\end{footnotesize}
\end{table}

Finite time averaging of the visibility function also results in loss of amplitude for objects not located at the phase-tracking centre. Nevertheless, at the low frequencies of interest in this work, this effect is much less significant than bandwidth smearing and is not used here as a directional filter.

\subsection{Measuring and correcting for DD phase errors  across the FoV}

LEAP uses standard phase self-calibration techniques and a local sky model  (i.e. a model of the calibrator) applied to phase rotated and frequency averaged datasets, to perform DD calibration in the directions of LEAP calibrators across the FoV.
At the MWA resolution the local sky model nearly always consists of a point-like source at the phase centre. 
When dealing with longer baselines, such as in the case of LOFAR, more complicated models including  high resolution structure may be required, but still over a small angular extent.
Note that since a global sky model is not required, each direction is an independent process and therefore all of them can be processed in parallel, regardless of the number of directions involved.
Hence this is an ``embarrassingly parallel'' problem and this is an important consideration for the data processing time.

The outcome of the phase self-calibration procedure is a set of time-variable antenna-based DDE corrections that are applied, in the visibility domain, to the original DI-calibrated datasets, and Fourier transformed to create an image, see Section \ref{sec:dd}. 
The DD calibration of a wide FoV is achieved by repeating the process described above in the direction of multiple LEAP calibrators within the FoV, sufficient to probe the ionospheric spatial structure, and with a solution time interval shorter than the ionospheric fluctuation timescales.  

\subsection{LEAP calibrators}
\label{subsec:leapcals}

A LEAP calibrator is a source that enables a precise and accurate measurement of the residual antenna-based phase errors  in that direction, remaining after DI calibration, on timescales smaller than the ionospheric fluctuations.
The requirements are those for a general purpose calibrator source, i.e. the accurate position of the source must be known, it should have compact structure or, if extended, the intensity distribution must be well known, and be sufficiently strong to have enough signal to noise ratio (SNR) for self-calibration techniques.
Additionally it must remain the dominant signal in the bandwidth smeared dataset.  
The latter is to ensure that the DDEs probe the distortions in the direction of the LEAP calibrator, which is achievable if the signal from the calibrator dominates over the noise contribution from the smeared sources in other directions;
in practise this depends on the interplay between the relative strength of the calibrator, the distribution of other sources in the field, the averaged bandwidth and the baseline lengths. 
Hence,  a LEAP calibrator has to fulfil two SNR thresholds. 
The second requirement sets a harder constraint, and results in a flux density threshold higher than that set by the raw sensitivity.
Section \ref{sec:cals} discusses the availability of LEAP calibrators for MWA observations, which results in a lower flux density limit of $\sim$2\,Jy.
The entries in a reference catalogue are a good starting point to select LEAP calibrators.
For example, the GLEAM catalogue has 13,288 sources with integrated flux densities larger than 2\,Jy at 84\,MHz, which corresponds to an average separation between these sources of 1.3\degr;
at 151 MHz there are 6341 sources and the average separation is 2.0\degr.


\subsection{Spatial and temporal characterisation of the ionospheric distortions}

The antenna-based DDE corrections from LEAP probe the ionospheric spatial distortions over the array, without any constraints on their functional dependence, in multiple directions.
As such they are sensitive to linear and higher order phase structure
at fine scales smaller than the size of the array, i.e. on physical scales $<$3\,km for MWA; 
and on any timescale, limited only by the correlator integration time and the instantaneous sensitivity. In comparison, image-based schemes are only sensitive to linear structures and the snapshot timescale.

The representation of the DDE values for a given direction with respect to the X-Y plane locations of MWA antennas on the ground defines an ionospheric screen, {\it hereafter} the DDE surface. 
Moreover, larger scale spatial variations can be explored by combining the measurements in different directions across the FoV. Such large scale studies have been performed \citep{loi_15,jordon_17} using image-based analysis, however these can not probe the fine scales.

\section{Observations and Methods}
\label{sec:obsmeth}
\subsection{Observations}
\label{sec:obs}

We have selected a subset of MWA GLEAM observations at two frequency bands: 72-103 and 139-170 MHz ({\it hereafter} 88 and 154\,MHz, respectively), to demonstrate the feasibility of LEAP for DD calibration.
The GLEAM survey strategy is described in detail in \citet{wayth_15}. Briefly, meridian drift scans were used to cover the entire sky visible to the MWA. 
The sky was divided into seven declination strips; a single declination setting was observed in a night. Within a night, the observations were conducted as a series of 2-min long scans at each frequency, cycling through all five frequency bands over 10 min. 

Our verification dataset comprises of observations at declination -13\degr, carried out on the night of 31$^{\rm st}$ October 2014 as part of the second year of GLEAM observations (Year-2). It consists of 7 sequential 2-min snapshots observed at each frequency, in an interleaving fashion,
spanning about 70 minutes. At both frequencies the pointing centres were in the range of 1.6--2.7h in Right Ascension (RA).
The common part of the sky within the wide FoV of MWA 
is our verification field, with a size of $\sim$15\degr\ across in RA centred on ca. 2.3h.
The subset of observations includes a range of weather conditions, significantly worsening towards the end, and allows a direct comparison of the calibration performance at two frequencies under very similar weather conditions.
In addition, we have used several other datasets to explore the LEAP performance under a wider range of weather regimes, including the most severe weather conditions during GLEAM Year-1 observations. These are observations at 88 MHz, where the effects are maximum.

\subsection{Analysis methods: DI-analysis}\label{sec:di}


The initial DI calibration of all the datasets presented in this paper is performed as described in \citet{nhw_gleam} for GLEAM Year-1 observations. Briefly,  
the \textsc{cotter} pre-processing pipeline \citep{offringa2015} is used to flag RFI, average the visibilities to a time resolution of 4\,s, a frequency resolution of 40\,kHz and convert the raw data to measurement sets of size $\sim$8GB.
The snapshots are calibrated in two stages:
an initial transfer of complex antenna-based gain solutions obtained from a calibrator observation and a self-calibration loop with hybrid mapping.
The calibrator used is Pictor A, observed on the same night and at the same frequencies as the target snapshots.
For each snapshot, a model for self-calibration is produced by imaging with a robust weighting factor \citep{briggs_95} of --1.0 (close to uniform weighting) down to the first negative CLEAN component using \textsc{wsclean} \citep{offringa2014}. 
%
This hybrid-mapping approach uses the initial snapshot image as the sky model for the bandpass phase and amplitude self-calibration, which
has the effect of absorbing the bulk ionospheric position shift, 
i.e. flux-weighted average of the apparent shifts of sources across the FoV,
into the snapshot images. 
This signature is therefore imprinted in the DI-calibrated visibility datasets.

After calibration each snapshot is imaged separately as above except that the entire image is CLEANed (with `joint-channel deconvolution' to account for the frequency span) to 3$\sigma$, a mask is constructed from the identified components and the CLEANing is continued with the mask to 1$\sigma$. This is conducted in an automated fashion using the `auto-mask' and `auto-threshold' parameters 
in \textsc{wsclean}. The snapshots are corrected using the primary beam model from \cite{sokolowski2017} and combined together  using the mosaicing software SWARP \citep{bertin_02}. 
These images have pixel sizes chosen such that the FWHM of the synthesised beam is sampled by at least four pixels; at 154 MHz, the pixel size is 32.7 arcsec and at 88 MHz, it is 57.4 arcsec. The snapshot images were 4000$\times$4000 pixels, which extends to 10 per cent of the peak primary beam response. The typical root mean square (rms) noise in the centre of the snapshot images, after deep cleaning, is 20 and 80\,mJy/beam at 154-MHz and 88-MHz, respectively.





\subsection{Analysis methods: LEAP DD-analysis}\label{sec:dd}

The first step of the DD analysis is to rotate  (using {\sc chgcentre}, part of the {\sc wsclean} package) 
the DI-calibrated visibility data to the positions of a selected number of LEAP calibrators within the FoV, 
followed by averaging over the total instantaneous bandwidth of 30.72 MHz (using CASA \citep{casa} task {\sc split}). 
This results in the same number of intermediate smaller datasets (of size $\sim$10\,MB), per snapshot, each with a LEAP calibrator at its phase centre. 
In practise, as part of the development of the method and empirical identification of calibrators, this was carried out for all MRC sources within the FoV with a lower flux density limit of 0.67\,Jy and that are morphological simple, following the same pre-selection criteria as in standard GLEAM ionospheric analysis \citep{nhw_gleam}.
The number of sources in each snapshot ranges between 100--1000, depending on the frequency and region of the sky. 
For the 53 datasets we studied for this paper we investigated 2,180 different candidate LEAP calibrator sources. Each of these were analysed, generating calibration solutions and a 10\degr$\times$10\degr\ image around the source. As we explored a range of solution parameters a total of 29,567 calibration sets and images were created.
The LEAP calibrators are the subset that satisfy the requirements in Section \ref{subsec:leapcals} and are presented in Section \ref{sec:cals}.

\begin{figure}
\centering
\includegraphics[width=0.7\textwidth]{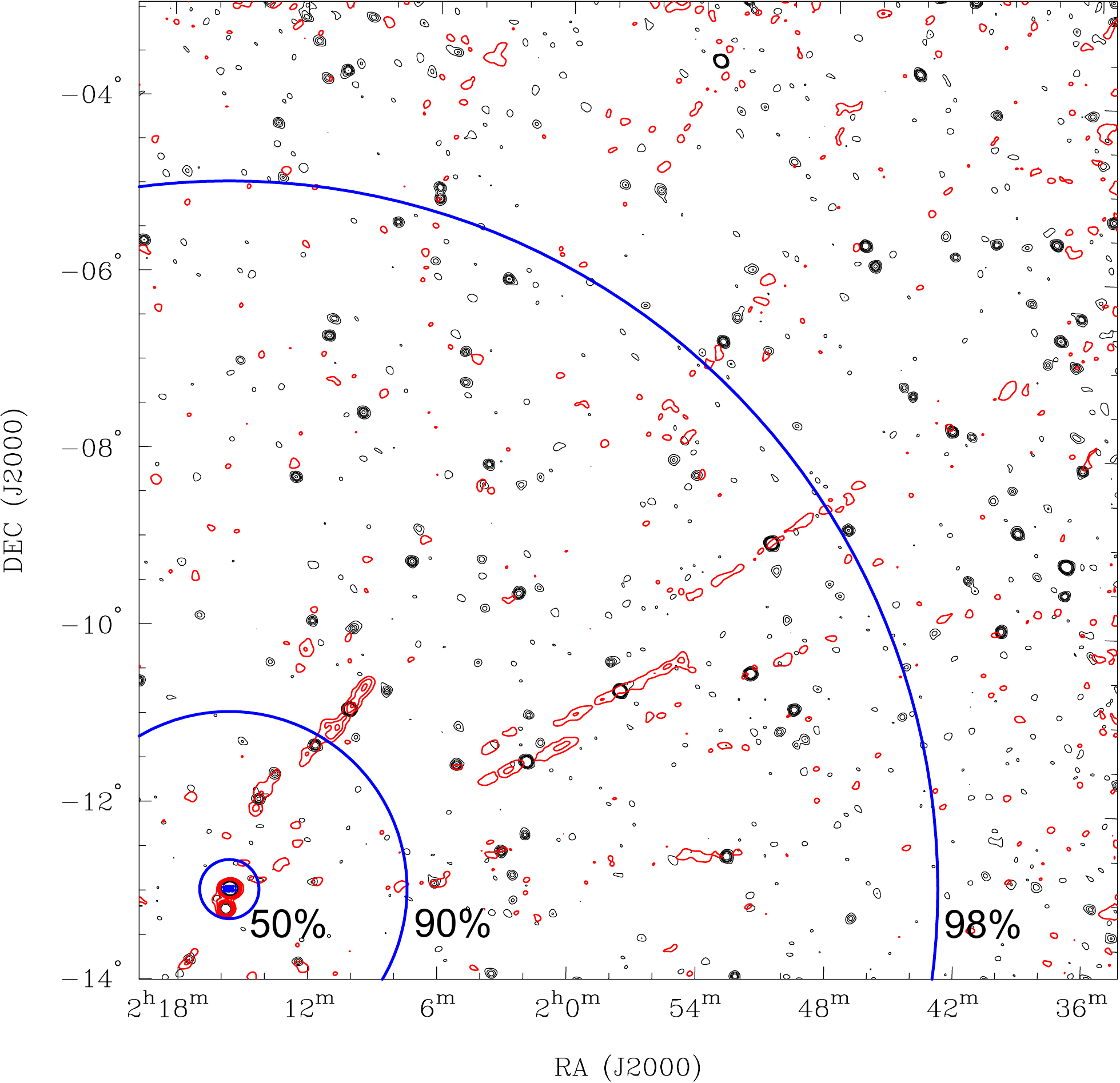}
\caption{Frequency averaging is an effective directional filter, to separate out the sky signal from directions of interest, for parallel processing. Plotted are contours (0.3, 0.6, 1.2 and
  1.8~Jy) for a 2 minute MWA snapshot image with a central frequency of 88 MHz and a bandwidth of 30.72 MHz, before (black) and after (red) frequency averaging. The source at the phase tracking-centre, whose apparent flux density is 15 Jy, is located in the bottom left corner and is unaffected by the averaging, whilst the many other sources in the sky are attenuated and smeared. Blue circles of $\sim$50\%, 90\% and 98\% attenuation ($\beta$=2, 12 and 48) are overlaid.
  \label{fig:maps_filter}}
\end{figure}

Fig. \ref{fig:maps_filter} 
shows an example of the efficiency of the bandwidth smearing effect for suppressing the sky signal 
away from a LEAP calibrator located at the phase centre of a snapshot image, for the frequency averaged snapshot dataset, superimposed on the image from the unaveraged dataset. 
This effective reduction of the FoV enables the calibration in a specific direction of interest using standard antenna-based phase self-calibration processes {against} a point source model (i.e. using {\sc gaincal}).
Baselines shorter than $\sim$400m are excluded to avoid contamination from remaining radially extended structure in the frequency-averaged image (in general, both smeared-out sources and diffuse galactic plane emission).
The antenna-based phase values output by {\sc gaincal} are, hereafter, referred to as the single-direction DDE set ({which also define the DDE surfaces}); they correct for all DD phase errors, such as ionospheric distortions and beam shape, 
in the direction of the LEAP calibrator.
The DD calibration of the whole FoV is obtained  by processing all the selected LEAP calibrators in parallel.

Our DD analysis considers two types of calibration solutions: 
on one hand, the single-direction DDE sets ({\it hereafter} full calibration); and also, for comparison, the values of the best fit plane to the DDE surfaces
({\it hereafter}  planar  calibration). 
The planar calibration leverages all the antenna solutions to increase the sensitivity. 
This is specially useful for the central antennas which can have a significant number of baselines shorter than the uv-range threshold, resulting in poor gain solutions. 
However it can not correct for fine scale higher order ionospheric distortions or antenna-beam phase imperfections. 
A planar surface corresponds to the first order of ionospheric disturbances, which is expected to be the dominant term and a valid approximation for most observations with Phase 1 MWA-scale arrays; it results in a refractive apparent source position shift in the image plane per solution interval. A variation of the full calibration adopts the planar solutions for the most central antennas. 
Note that both calibrations originate from the same run of {\sc gaincal}, for each direction of interest.

Each set of calibration solutions 
are applied (using task {\sc applycal}) to the original (i.e. unaveraged and unrotated) DI-calibrated dataset,  which then is imaged to produce multiple snapshot images each with a region of improvement, 
{\it hereafter} single-direction snapshot images.  
The imaging was done with {\sc wsclean}, as described for the final stage of DI-analysis; note that for best performance the imaging of the full FoV is required to deconvolve all visible sources and reduce the noise in the region around the LEAP calibrator. 

For each of the calibrators within the verification field, the corresponding 7 single-direction snapshot images from the verification dataset are co-added; this results in single-direction co-added images.
Both types of images are used for the characterisation study of the LEAP scheme at 88 and 154\,MHz, by comparison with the corresponding DI-calibrated images (see Sections \ref{sec:fom1} and \ref{sec:fom2}); this provides empirical estimates of the size of the regions  with improved calibration (the ``facets'').
Finally, we produce a DD-mosaiced image of the verification field, by selecting and combining the facets from the single-direction co-added images for a subsample of calibrators, sufficient to be an improvement over the DI-calibration; this is described in  Section \ref{sec:mosaic}. 
Furthermore, the LEAP calibrators are directly tied to the astrometric frame, hence the facet images will be accurately registered.

\section{Results and Discussions}\label{sec:res}

We present the summary of the results from the comparative analysis of the verification and additional Year-1 datasets, at 154 and 88 MHz.
This comprises i) the characterisation and feasibility of the LEAP calibration scheme, ii) the reconstruction of an example of a multi-direction DD-mosaiced image, iii) characterisation of the fine scale ionospheric distortions and defocusing effects, iv) an error analysis and v) assessment of calibrator availability.

\subsection{LEAP performance evaluation studies}\label{sec:e2e}

The aim of these studies is to characterise the performance of the LEAP scheme in the direction of the calibrator and in the region around it.
We examined the 25 LEAP calibrators in the verification field along with the strongest sources within an angular distance of $\sim$6\degr\ from each LEAP calibrator. 
The flux densities of the calibrators range from 20\,Jy  (\pone) to 2\,Jy (\ptwo), at 154\,MHz.
%
We have selected two figures of merit (FOM) to evaluate the effectiveness of the ionospheric calibration based on the signature of errors in the 
images,  which are: 
i) The source position jitter for the 50 strongest sources, across the 7 snapshot-images spanning $\sim$70 minutes (FOM-1), 
and ii) the source peak intensities for the 100 strongest sources, measured from the co-added images (FOM-2).

Here we present the findings for the strongest and weakest calibrators, and for a representative of the intermediate flux densities: \pfive, with a flux density of 4\,Jy at 154\,MHz.

The DD analysis is comprised of  two types of calibration solutions: 
full calibration, from the unconstrained antenna-based DDEs,
and the planar calibration, which is a fit to those, 
both using a solution interval of 120\,s.
We have also investigated using an interval of 30\,s and 
the results of the analysis showed no significant differences for the verification dataset, which was observed under moderate weather conditions.

\subsubsection{FOM-1: Source position jitter}\label{sec:fom1}

\begin{figure}
\centering
\includegraphics[width=0.4\textwidth]{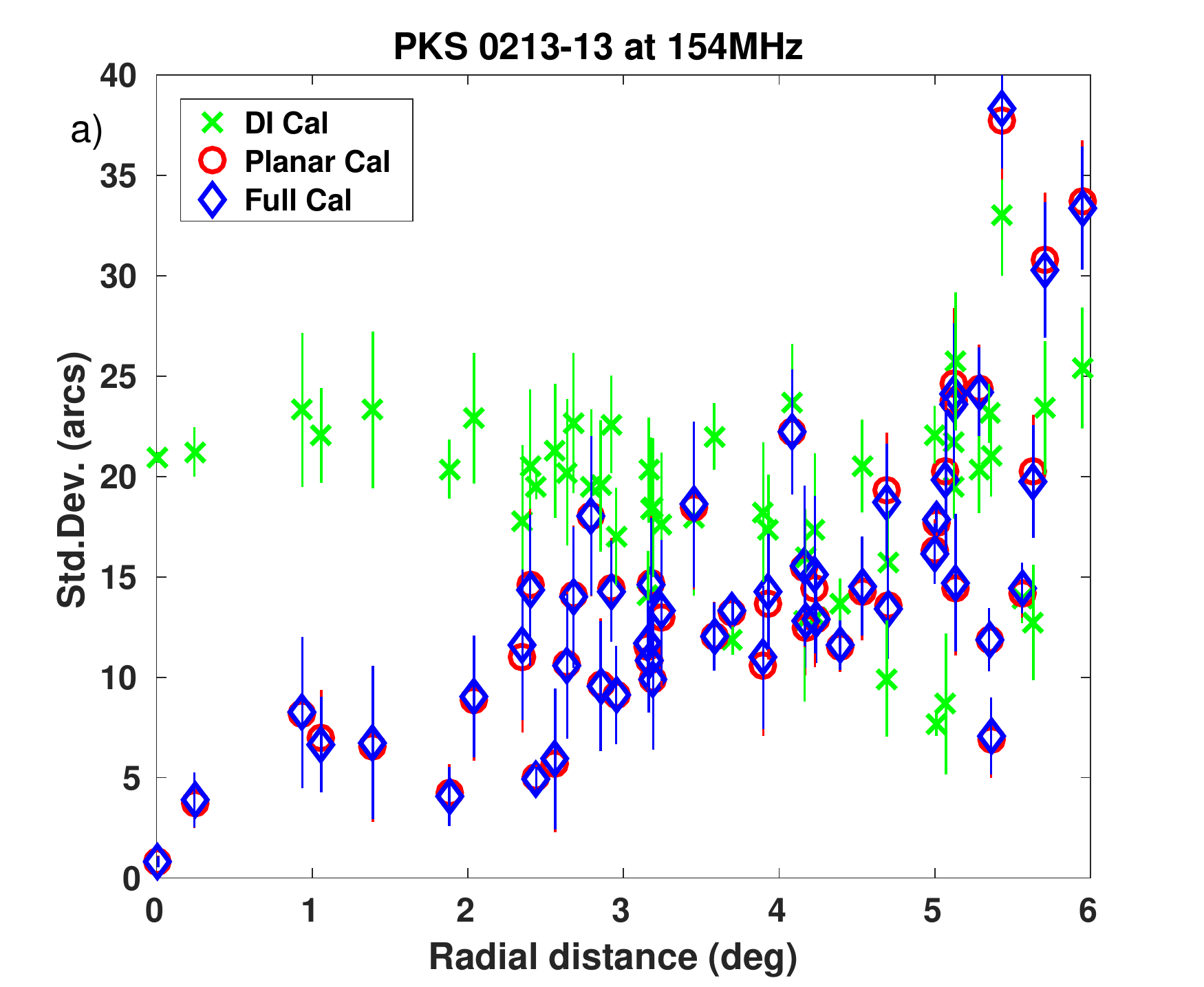}
\includegraphics[width=0.4\textwidth]{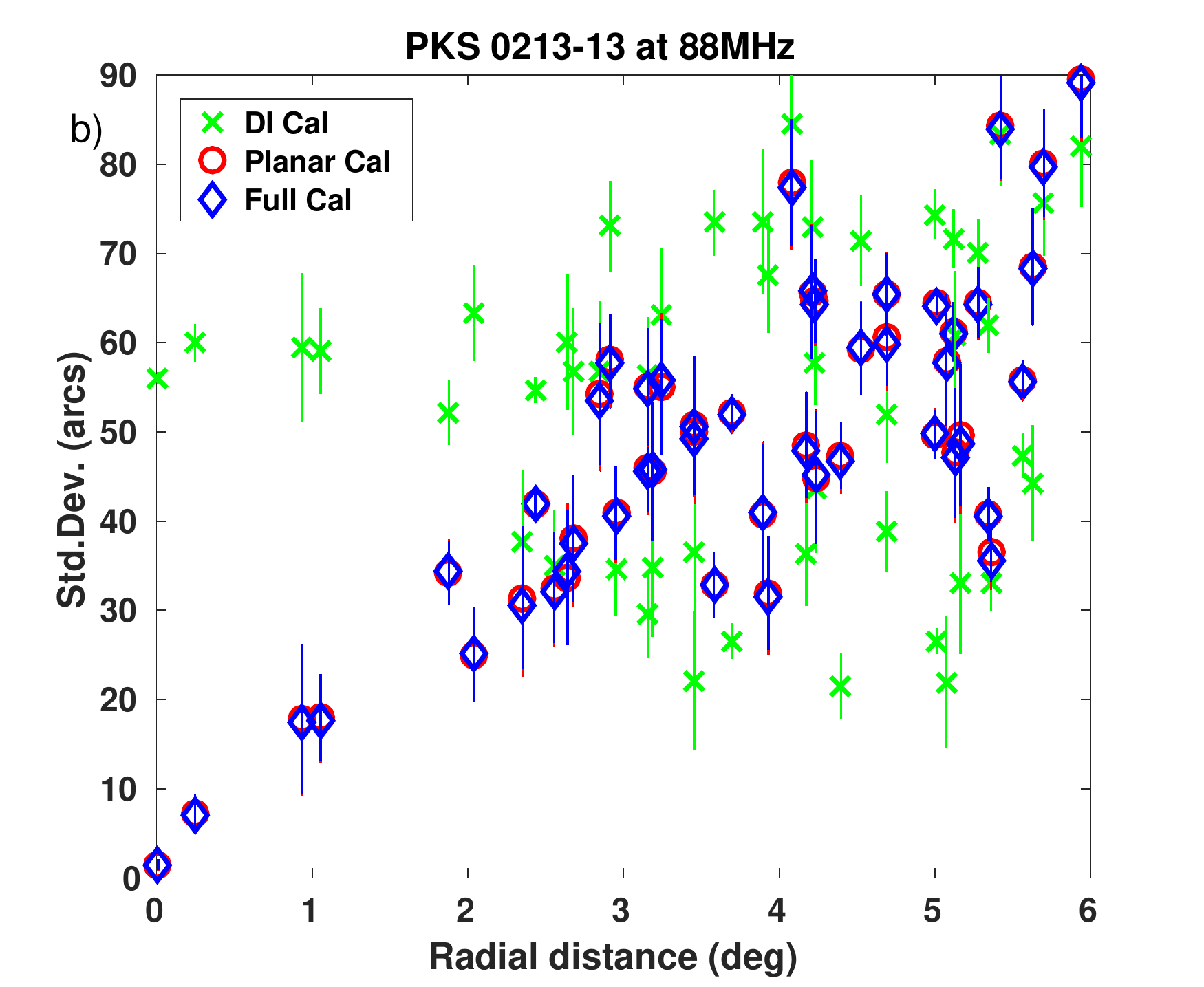}
\includegraphics[width=0.4\textwidth]{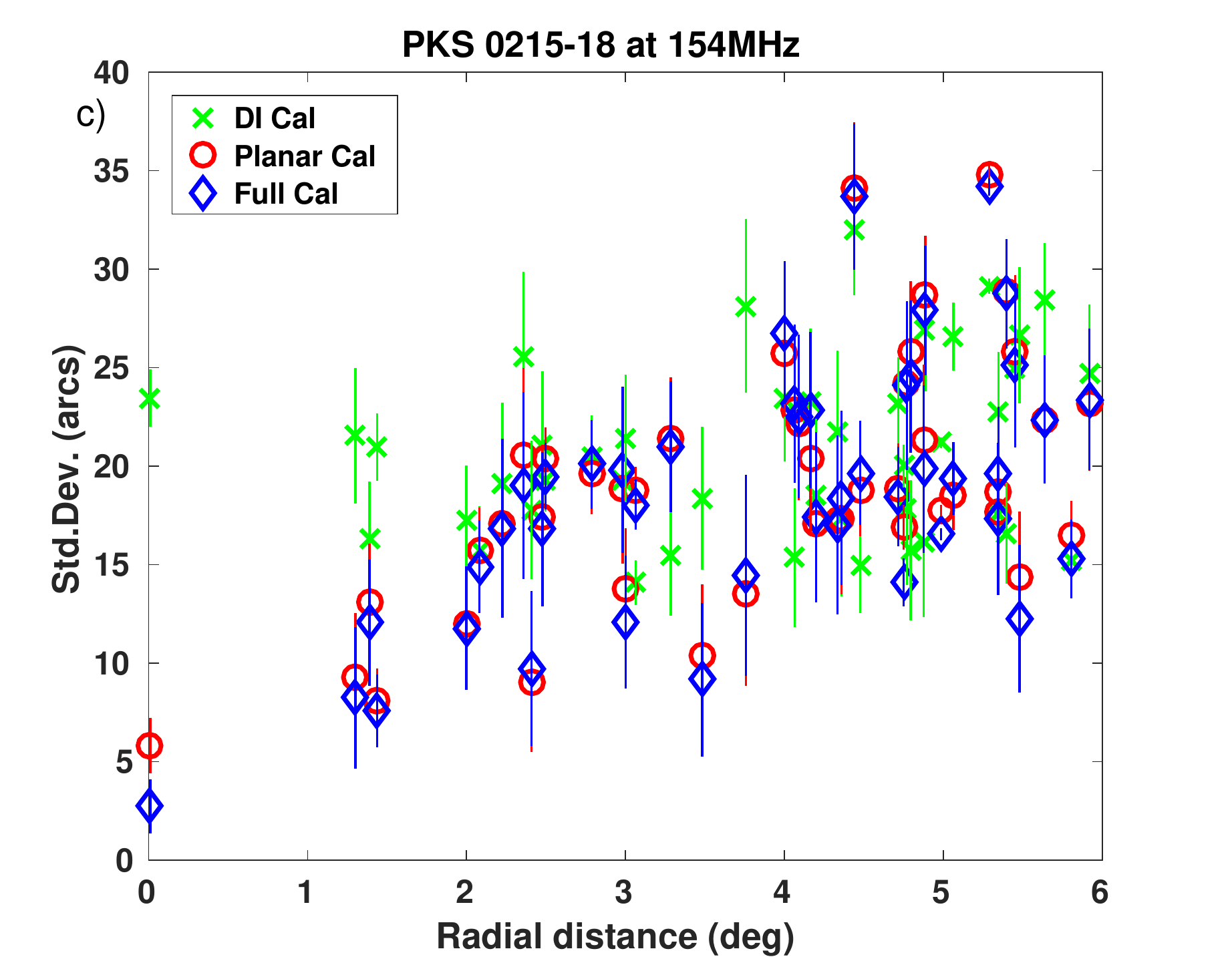}
\includegraphics[width=0.4\textwidth]{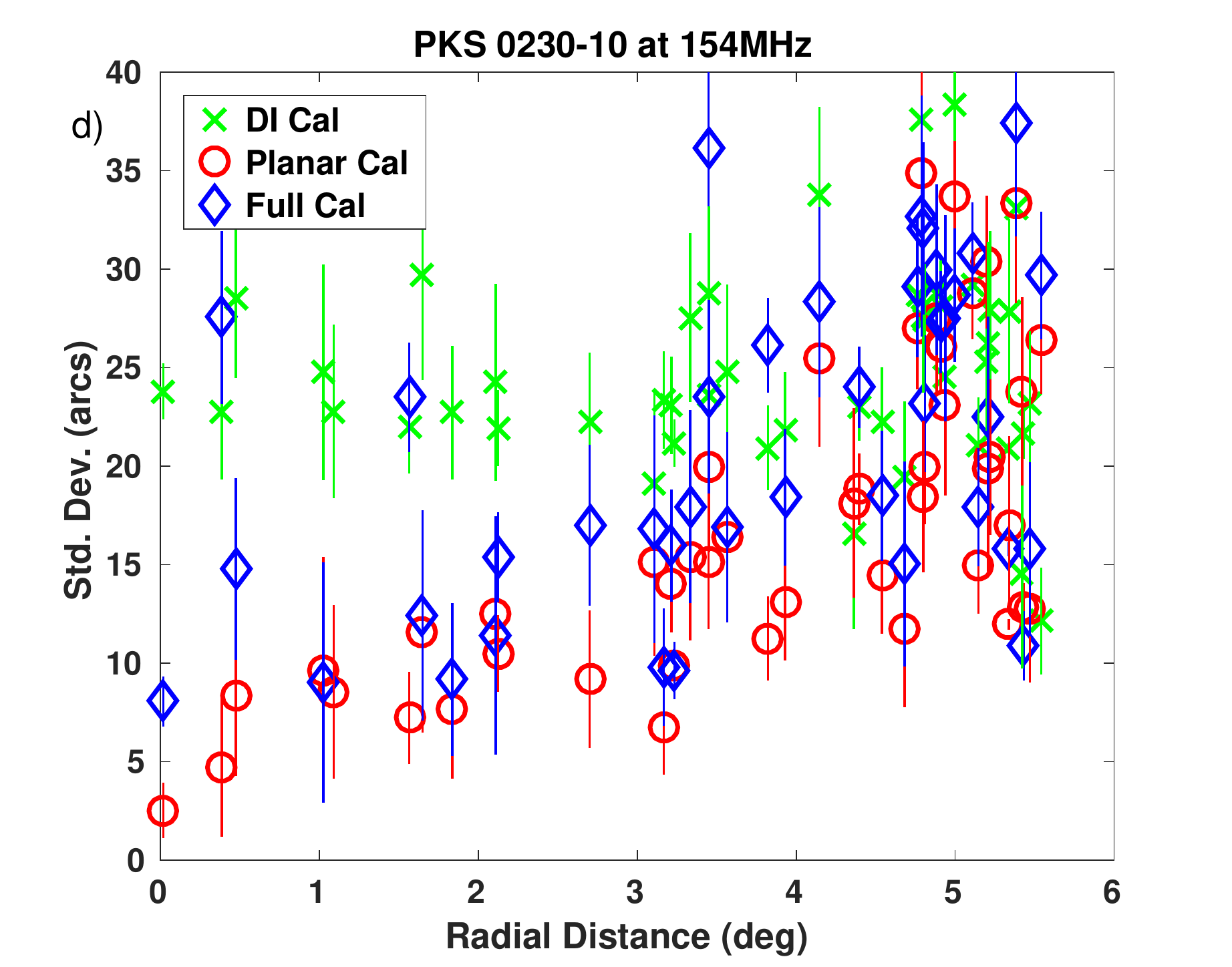}
\caption{Performance Evaluation Studies (FOM-1): Standard deviation of observed source positions over the seven snapshot images, as a function of angular separation from the LEAP calibrator, for a strong source (\pone) at {\it a)} 154\, MHz and {\it b)} 88\, MHz. The results from the DD-calibrated data are plotted with blue diamonds and red circles, for full and planar calibration, respectively. The results for the DI-calibrated data are plotted with green crosses. $\pm1\sigma$ thermal errors  are indicated with vertical lines.
The outcome of the full and planar calibrations are identical and improve the source stability with respect to the DI-calibration up to an angular separation of $\sim$3--5\degr\ at 154 and  88\,MHz. 
Figure {\it c)} is for an intermediate flux source (\pfive), 
whereas {\it d)} is for a weak source (\ptwo), both at 154\,MHz. For the weak case the planar calibration gives better solutions than the full calibration for some cases, whilst for the intermediate case the results are almost indistinguishable. In both cases the DD-results intercept the DI-results at a similar range of angular separations as for the strong source case.
\label{fig:fom1}
}
\end{figure}

We use the stability of source positions (i.e. position jitter) 
across the 7 DD 
single-direction snapshot images as a FOM and compare it with the outcomes from DI-calibration. 
Fig. \ref{fig:fom1}a  shows the standard deviations versus the angular separation from the strong calibrator source for DI- and DD-analysis, at 154\,MHz. The intrinsic thermal error bars for each source, given by the ratio of the beam size over SNR \citep{condon_98}, show no correlation with separation, indicating that systematic contributions are the dominant error.
Fig. \ref{fig:fom1}b is the equivalent to Fig. \ref{fig:fom1}a for the interleaving observations at 88\,MHz.
Both figures display similar qualitative trends: 
for the DD-calibration, the position stability is  highest in the direction of the calibrator (reaching the thermal noise level, as would be expected for self-calibration) 
and degrades for sources at increasing angular distance from the calibrator.
It reaches the level of the DI-calibration stability (which, in this case, is approximately a constant plateau) at an angular separation of about 3--5\degr. 
In absolute values, the standard deviations at 88 MHz are more significant, but the intercept is at a similar angular separation. The full and planar calibration result in practically identical improvements. 

These findings are in agreement with expectations from the functional dependence of astrometric errors, that is $\sigma_{pos} \propto \Delta{\rm TEC_{DD}} / \nu^{2}$, for observations at two frequencies under the same underlying ionospheric conditions, where $\Delta{\rm TEC_{DD}}$ is the difference in $\Delta$TEC observed for the two directions and increases with angular separation from the calibrator.
The same frequency dependence is seen in the astrometric errors from the DI-analysis at both frequencies, which are 20 and 60\,arcsec (in the region of the LEAP calibrator), in agreement with $(154 {\rm MHz}/88 {\rm MHz})^2 \sim 3.1$;  
these errors correspond to residual differential ionospheric errors of $\Delta{\rm TEC_{DD}} \sim$0.02 TECU over the array
(which result in phase errors of  $\sim$60\degr\ at 154\,MHz, for the longest baselines).

Figs. \ref{fig:fom1}c and d show the position stability versus angular separation for the intermediate and weakest calibrator sources, respectively, at 154\,MHz.
The general trends described above for the strong calibrator continue to apply here, with the proviso that for the weakest calibrator  the planar calibration is significantly better than the full calibration in some cases.
This is in agreement with the expectations for increased sensitivity when fitting a reduced number of parameters. 
The intercept of the DI and DD results are at a similar range of angular separations as for the strong source case. This gives the  angular radius of the region of improved calibration
and matches the 10\degr\ smoothing diameter used in the standard GLEAM image warping for DD-corrections.

\subsubsection{FOM-2: Source peak intensities}\label{sec:fom2}

\begin{figure}
\centering
\includegraphics[angle=0,width=0.49\textwidth]{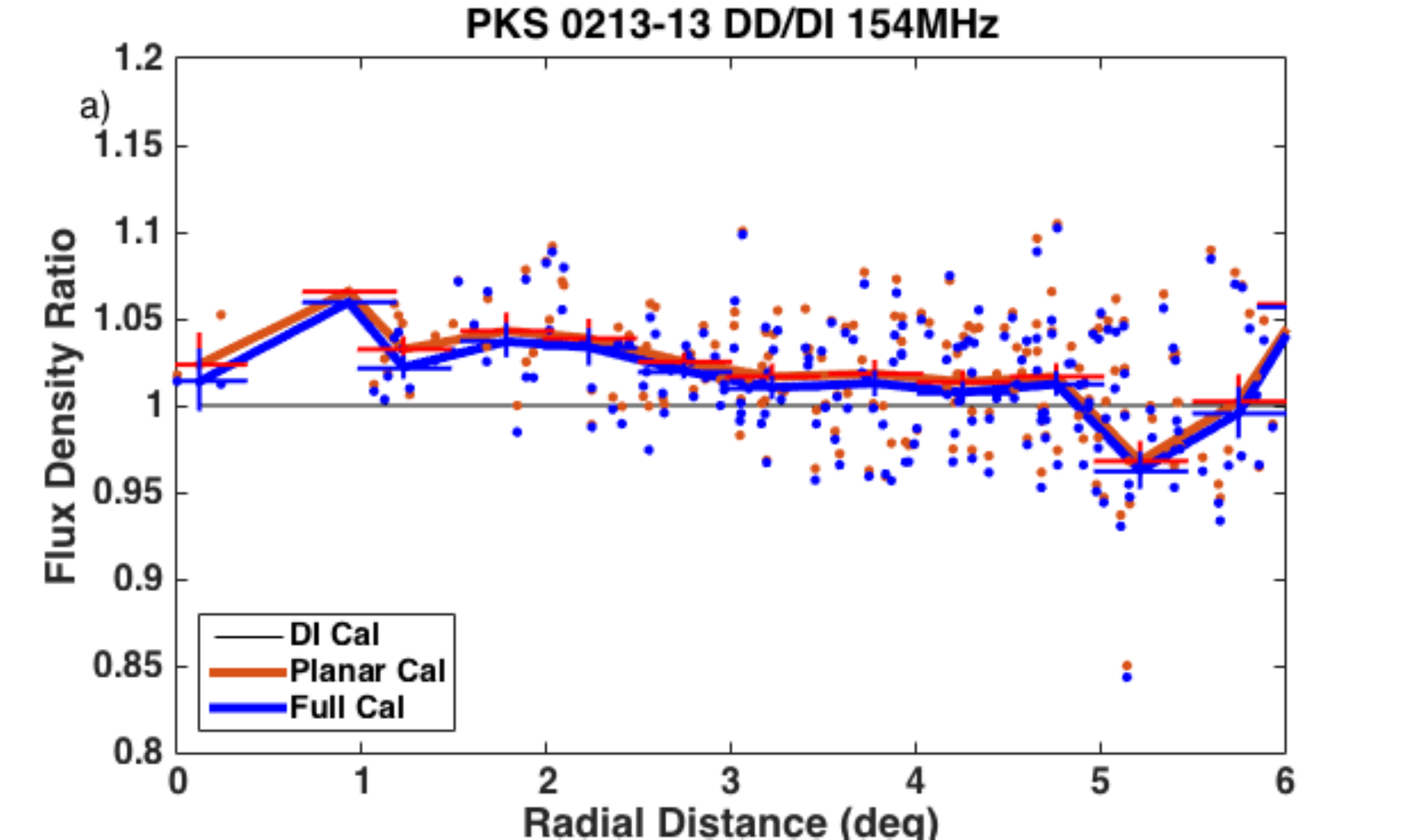}
\includegraphics[angle=0,width=0.49\textwidth]{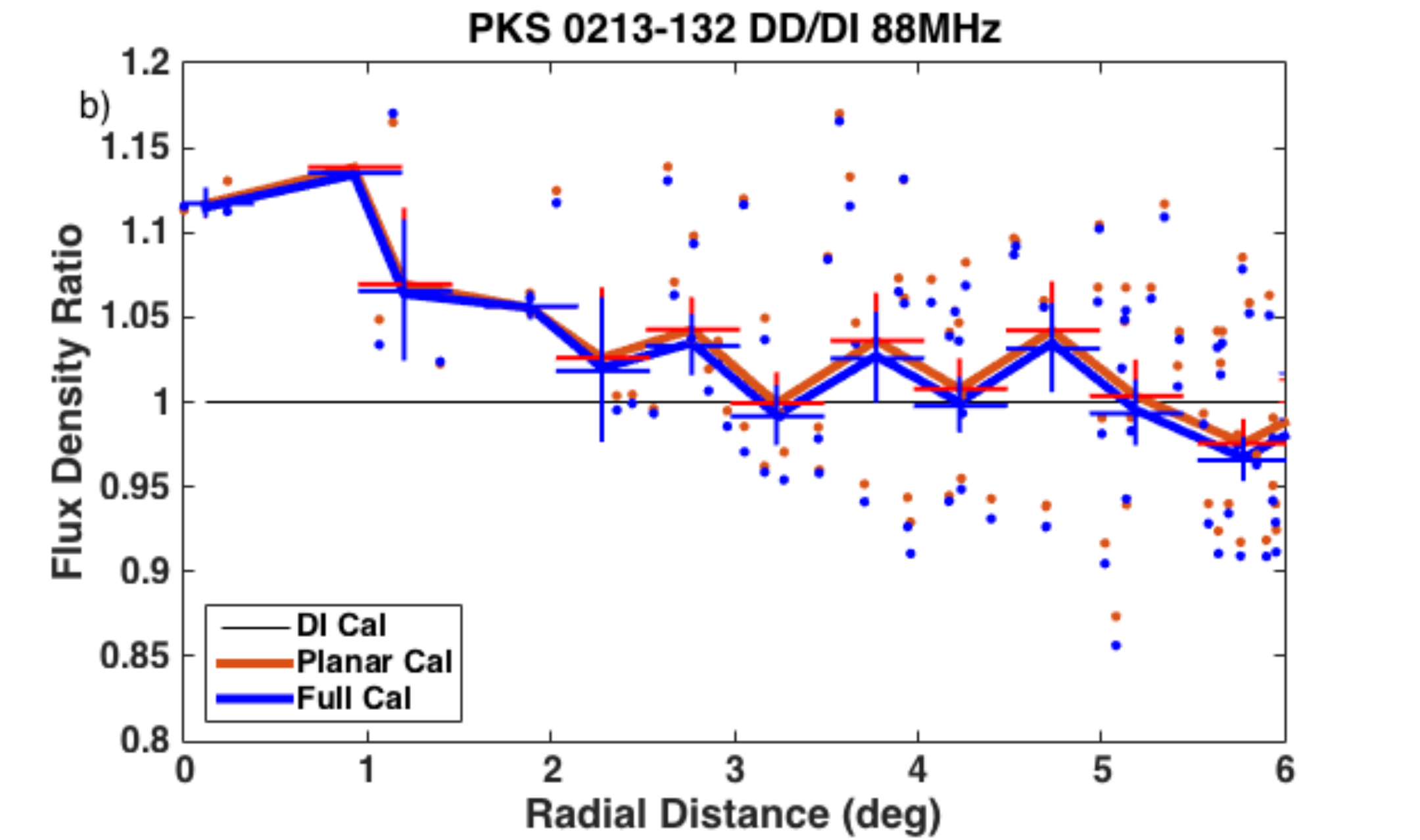}
\includegraphics[angle=0,width=0.49\textwidth]{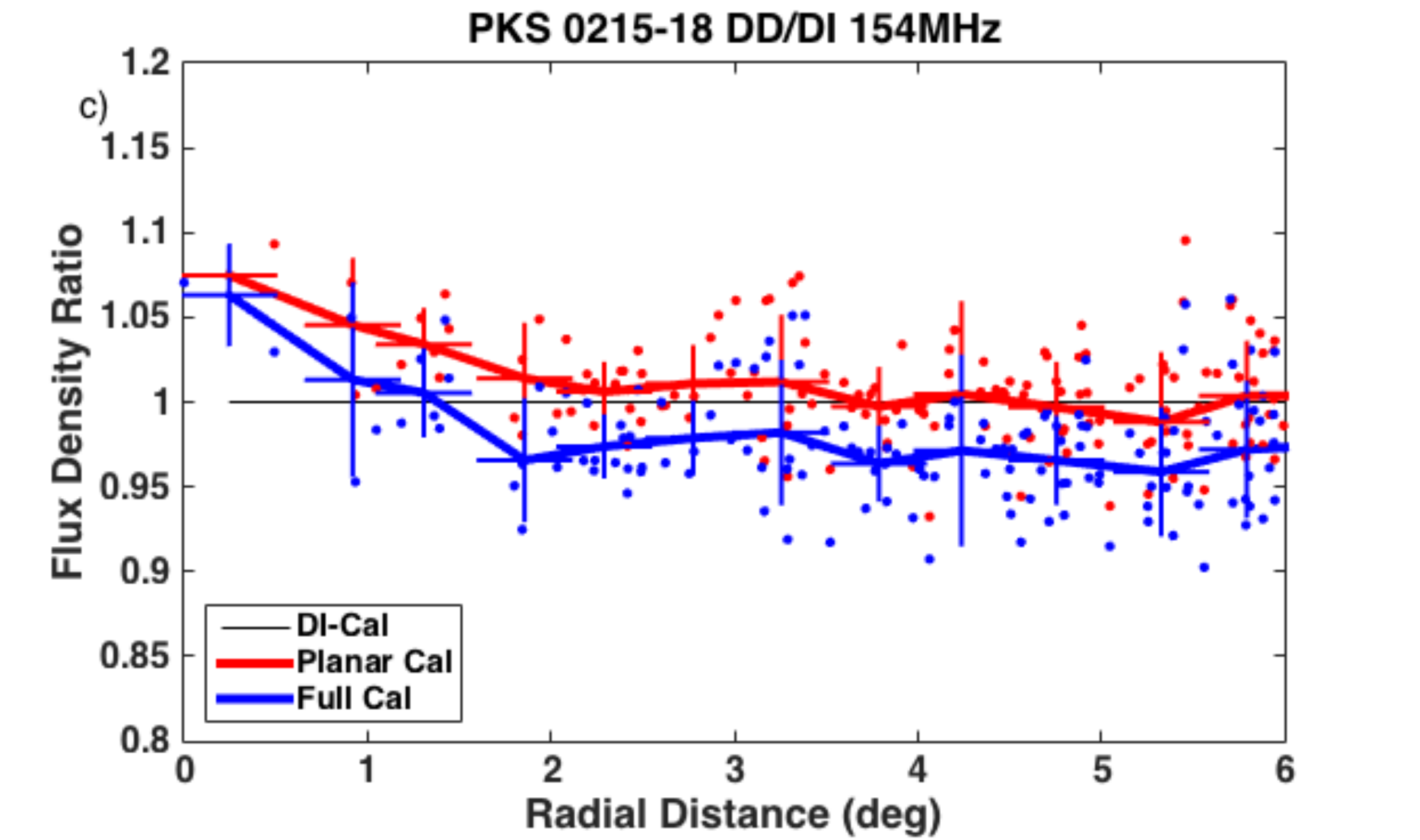}
\includegraphics[angle=0,width=0.49\textwidth]{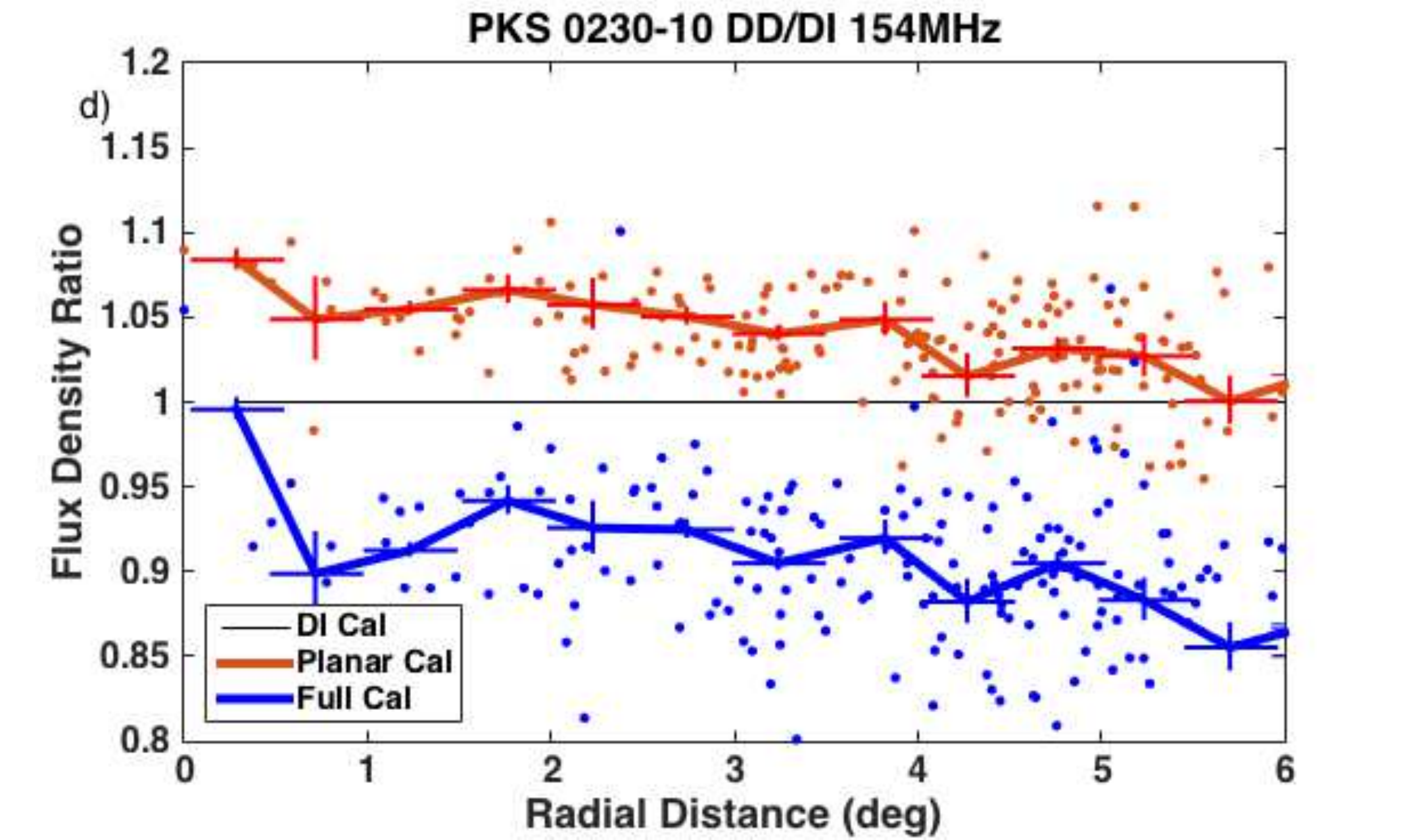}
\caption{Performance Evaluation Studies (FOM-2): The ratio of the source flux densities in the single-direction co-added images to that in the DI co-added image, against the angular separation from the LEAP calibrator, for the 100 strongest sources within a radial distance of $6\degr$.
Blue dots are for DD full calibration, red dots are for DD planar calibration; the continuous lines are the 0.5\degr\ binned average values, with the angular span of the bin and the 1$\sigma$ error indicated with the light crosses. 
Figure {\it a}) is for a strong calibrator (\pone) at 154\,MHz, where the full and the planar calibration result in similar improvements with respect to the DI calibration, and {\it b}) similarly for 88\,MHz. 
Figure {\it c)} is for an intermediate flux density source (\pfive), 
whereas {\it d)} is for a weak source (\ptwo), both at 154\,MHz. For the weak case the full calibration clearly gives worse solutions than the DI calibration, whilst for the intermediate case the full calibration is better than the DI calibration up to a separation of $\sim$1.5\degr. In both cases the planar calibration intercepts the DI-results at a similar range of angular separations as for the strong source case.
%
\label{fig:fom2}}
\end{figure}

We used the ratio of the source peak intensities in the DD single-direction co-added over the DI-  co-added images as a second FOM.
Fig. \ref{fig:fom2}a shows the ratio of intensities 
versus the angular separation from the strong calibrator  source  
at 154 MHz. These results correspond to the case shown in Fig. \ref{fig:fom1}a for the position stability;
Fig. \ref{fig:fom2}b shows the results from the analysis at 88\,MHz corresponding to the case in Fig. \ref{fig:fom1}b. 
Figs. \ref{fig:fom2}c and d are for the region around the intermediate and weakest flux density calibrator sources, respectively, which correspond to Figs. \ref{fig:fom1}c and d.

For the strong calibrator, both full and planar calibration solutions result in a similar improvement and over a similar angular range as for FOM-1 with respect to DI-calibration. 
For the intermediate flux density case, both the full and planar calibration improve the DI-calibration over a region around the calibrator, with the latter extending across a larger angular range similar to that in FOM-1.
The full calibration provides improved DD calibration out to $\sim$1.5\degr.
Fig \ref{fig:fom2}d, for the weakest source, shows the limiting case. Here the full calibration is never better than the DI calibration, however the planar calibration provides an improved DD correction across a similar angular range as FOM-1. 

\subsubsection{Implications of FOM-1 and FOM-2 results}

The FOM-2 indicates that the performance of the full calibration gets progressively worse with decreasing calibrator flux density. 
The lower flux density combined with the uvrange selection in gaincal 
results in much less signal contributing to the solutions for the central antennas that predominantly form the (excluded) shorter baselines. 
This results in reduced peak intensity in the image, compared to planar calibration. 



However, the FOM-1 indicates that the source position stability from full and planar calibration shows similar improvements with respect to DI-calibration for the strong and intermediate case and even in most cases for the weakest source
(see Fig. \ref{fig:fom1}). In other words, astrometry, which is dominated by the longest baselines, 
is more robust against the calibration issues related to the weaker sources.

For a MWA Phase 1 type array, limited to short baselines only,  with a ``compact core'' of closely spaced array elements, moderate sensitivity and 30 MHz maximum instantaneous bandwidth,  
the planar calibration has the best general performance, with a larger number of calibrators available and valid for the majority of weather conditions, as shown in Section \ref{sec:screen}.
It retains the astrometric correction, and provides good solutions for the central stations under the assumption of first order ionospheric structure over the array.
Nevertheless, full calibration is required in some cases 
i.e. with poor weather or when the antenna-beams are not identical, requiring a higher minimum calibrator flux density 
threshold and/or reduced facet sizes.

The existing MWA Phase 2 upgraded array has double the diameter and expands the number of long baselines compared to Phase 1.
This will address the (weak source) limitations of full calibration for MWA Phase 1. 
Furthermore, full calibration is equipped to deal with the higher order ionospheric structures that will become significant due to the longer baselines.
With the increased sensitivity for the planned instruments, MWA Phase 3 and SKA-Low, such effects will also be significant even with the short baselines.
It should be noted that the planned larger instantaneous bandwidth will also contribute to improving the full calibration performance.
 Therefore the next generation of instruments will both enable and require full calibration.

\subsection{DD-mosaiced images}\label{sec:mosaic}

\begin{figure}
\centering
\includegraphics[width=0.7\textwidth,angle=0]{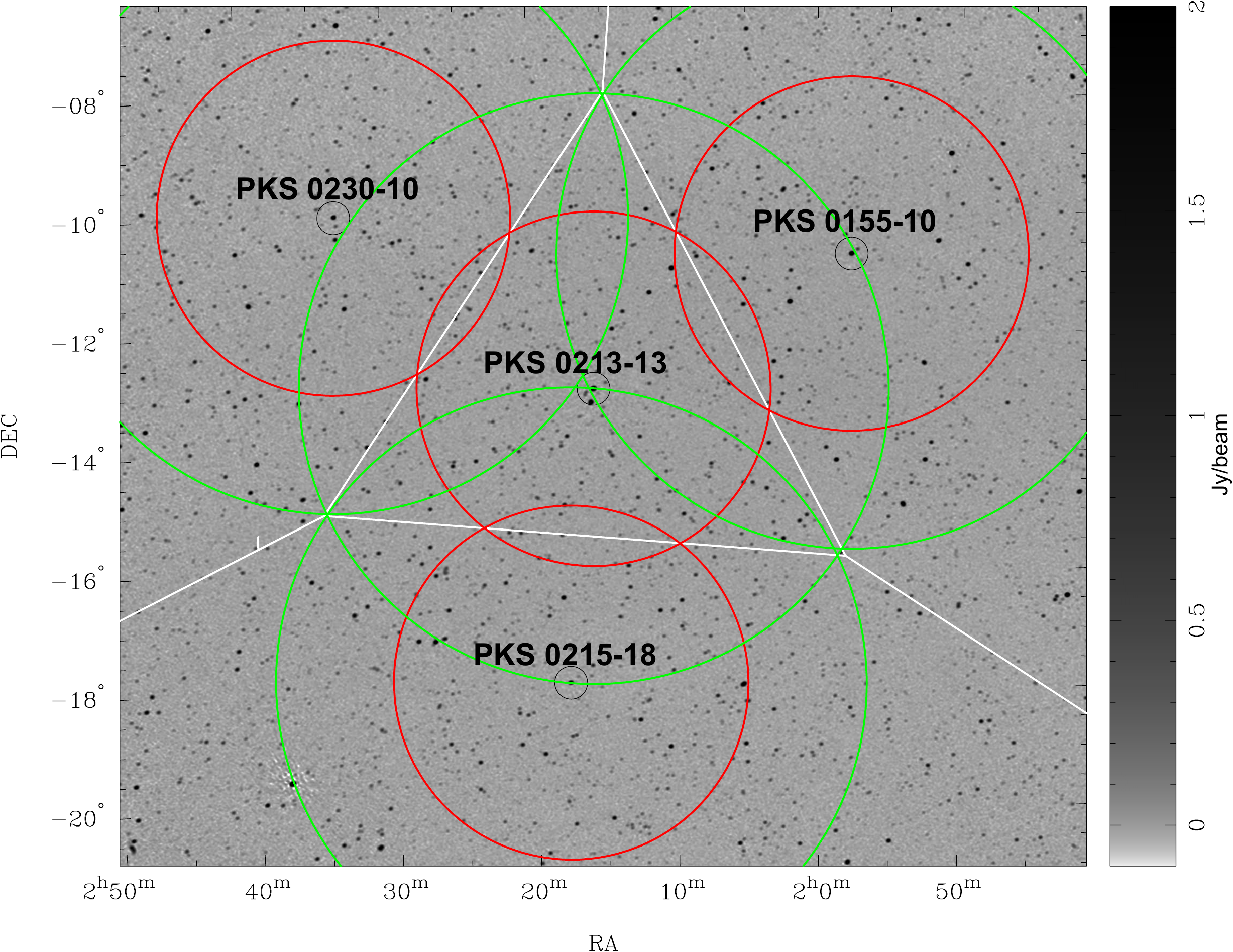}
\caption{The multi-direction DD-mosaiced image of the verification field at 154 MHz, using single-direction DDE corrections from the four LEAP calibrators, ringed and labelled with their PKS catalogue name. The regions of calibration validity around each calibrator are marked with red circles of radius 3\degr\ and green circles of 5\degr.
The greyscale is logarithmic, and runs between -0.1 to 2\,Jy/beam (-1 to 20\,$\sigma$). Also indicated are the facet regions around these sources, marked by white solid lines. Each facet was imaged independently with  single-direction co-added images from the 7 snapshots and merged together at the facet boundaries.\label{fig:mosaic}}
\end{figure}

DD-calibrated large FoV images are most simply achieved by masking and combining the regions of improved quality around each calibrator across the FoV,
so called facets, as demonstrated for LOFAR \citep{vanweeren_16}.
We follow a similar approach here to image the verification field 
using the planar calibration derived from the LEAP scheme at 154 MHz.
In this demonstration we selected a distribution of four LEAP calibrators (from 25 in the verification field) that minimises the number of facets required to achieve a net improvement with respect to DI calibration across most of the 15\degr$\times$15\degr\ field, 
with the outer calibrators within 6\degr\ angular separation from the central source, as shown in Fig. \ref{fig:mosaic}. 
Three of the calibrators correspond to the examples shown in Sections \ref{sec:fom1} and \ref{sec:fom2}.
The straight lines are the borders between the facets containing the sources that are closest to its calibrator; the large circles around the position of each calibrator show the regions within 3 and 5\degr\ angular separation.
These areas of validity are in agreement with the results derived from the FOM studies.
We created four single-direction co-added images each combining the 7 single-direction snapshot images from the analysis of the verification dataset, using SWARP as described in Section \ref{sec:di}, in the direction of the four LEAP calibrators. 
Finally, from each of the co-added images we extracted the corresponding facet region with improved calibration, 
and form the final DD-mosaiced image, shown in Fig. \ref{fig:mosaic}. 

The improvements in this image, and the level of precision achieved, are quantified in the FOM studies (Section \ref{sec:fom1} and \ref{sec:fom2}). Additionally, these studies also quantify the further improvements expected from reducing the size of the facets.
This can be achieved by: i) increasing the number of LEAP calibrators used in the analysis, and/or ii) adding ``virtual calibrators'' by interpolating the single-direction DDEs values to infer the calibration solutions at intermediate positions. 

For absolute astrometric validation we compare the positions of sources in the image calibrated with LEAP, shown in Fig. \ref{fig:mosaic}, with the positions listed in the NVSS catalogue. 
We used a simple cross-matching criteria of being within 20\arcsec\ of the expected position.
We calculated the flux-weighted mean and standard deviation of the differences between the positions, in the area of improved calibration (i.e. within 5\degr\ of a LEAP calibrator) to be -0.4$\pm$4.7\,arcsec, -0.9$\pm$2.5\,arcsec in RA and declination, respectively. 
To compare the visibility-based and the image-based DD-calibration methods we would carry out the same calculation for the common part of the GLEAM image made with the standard analysis, using the same observations.
However, we are unable to compare our image with that from Year-2 standard analysis, as these are not yet released (Franzen et al., in prep).
Therefore we have used the image from Year-1 observations \citep{nhw_gleam}. For this the flux-weighted mean and standard deviation are 
0.1$\pm$3.6\,arcsec, 1.8$\pm$3.8\,arcsec, in RA and declination, respectively.
%
We conclude that the LEAP image, from Year-2 data, and the standard analysis image, from Year-1 data, show no significant differences. 
The total errors are the same, but LEAP is slightly better in declination and worse in RA. We note that these observations were performed under different weather conditions and multiple declination strips contributed to the Year-1 image, making a decisive comparison difficult.
The good agreement between the two calibration schemes comes about as, for the Phase 1 of MWA, the ionospheric phase screens are approximately planar, under most weather conditions.


The LEAP calibration pipeline (i.e. copying of the measurement set, phase rotation, frequency averaging and self-calibration) only takes a few minutes using one node of our small departmental cluster. Many parallel processes could be sustained with no loss of performance, until we reached the I/O bottleneck on the node local disks.
Nevertheless, the imaging time scales linearly with the number of facets in the current imaging pipeline, although in principle these also could run in parallel. This bottle-neck is why we have limited ourselves to four facets in this demonstration. 

Alternative imaging approaches are possible. 
{\sc ddfacet} \citep{ddf_tasse}, for example, does not require additional computational resources for imaging with or without DD corrections, therefore adding additional DD corrections does not impact on the compute-time. 
We have successfully imaged the multiple MWA snapshots, with multiple DD LEAP calibration solutions applied, in a single pass of {\sc ddfacet}. However this was performed without applying true MWA beam corrections, which we are in the process of implementing for future use.
{\sc ddfacet} has other benefits, which will become more significant as the baselines get longer and the uv-coverage becomes more sparse. Sources distant from the LEAP calibrator are expected to be poorly calibrated. Currently, for the small footprint of MWA, the ionospheric errors propagate into apparent position shifts which are absorbed into the local sky-model. Therefore they do not contaminate the areas that have been well-calibrated. For LOFAR-scale arrays these conditions do not hold and residual errors cause source defocusing and contaminate large areas; this might be a consideration also with  MWA Phase 2. 
Such effects will strengthen the demand for the use of {\sc ddfacet} as the imager of choice. 

\subsection{Ionospheric phase screens and image defocusing}\label{sec:screen}

A byproduct of the LEAP scheme is a vast dataset for ionospheric studies. 
The DDE surfaces, or ionospheric screens, provide an instantaneous imprint of the fine scale ($<$3km) spatial phase structure of the ionospheric distortions 
over the array, in the directions of the LEAP calibrators within the FoV.

\begin{figure}
\centering

\includegraphics[width=0.5\textwidth]{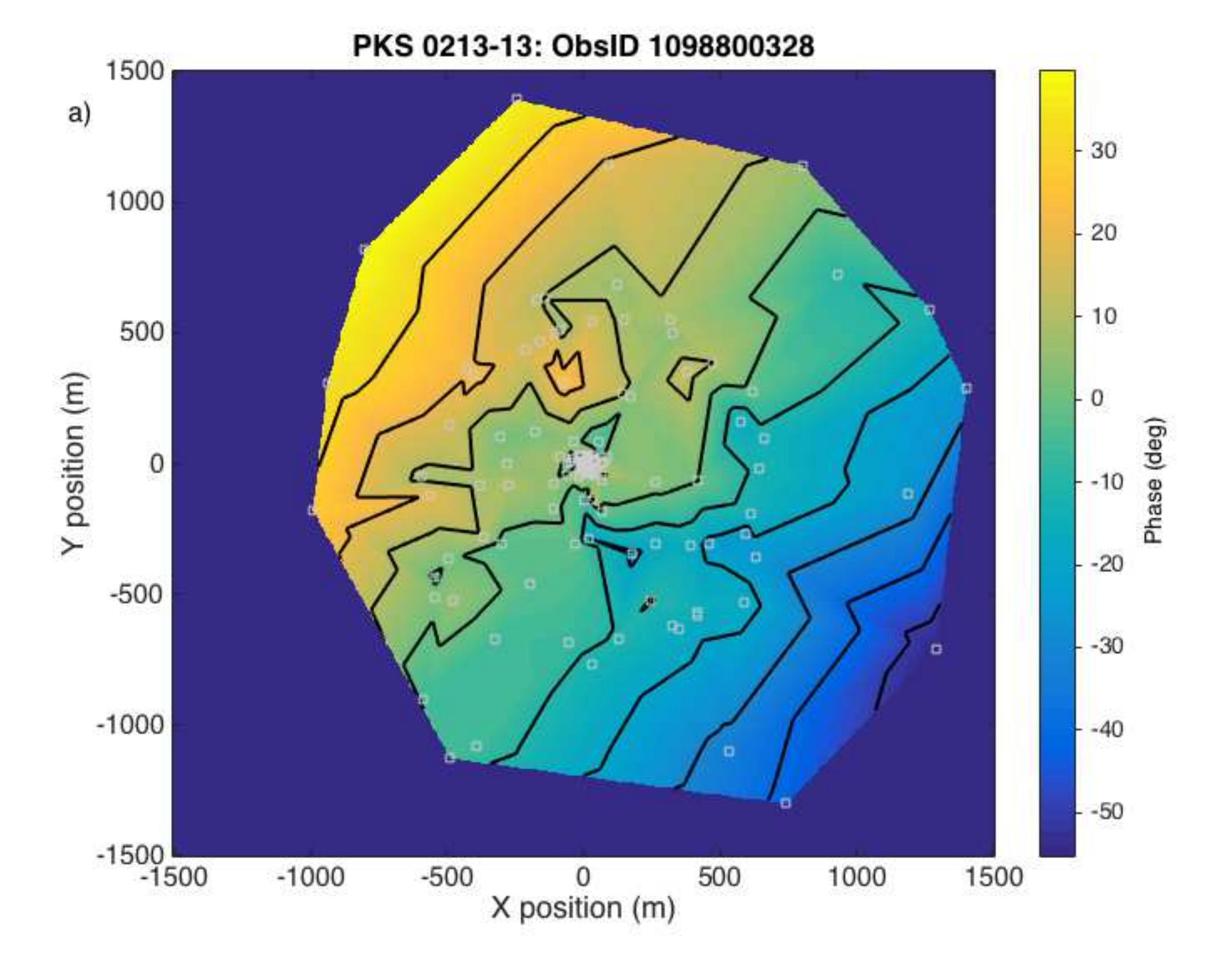}
\includegraphics[width=0.5\textwidth]{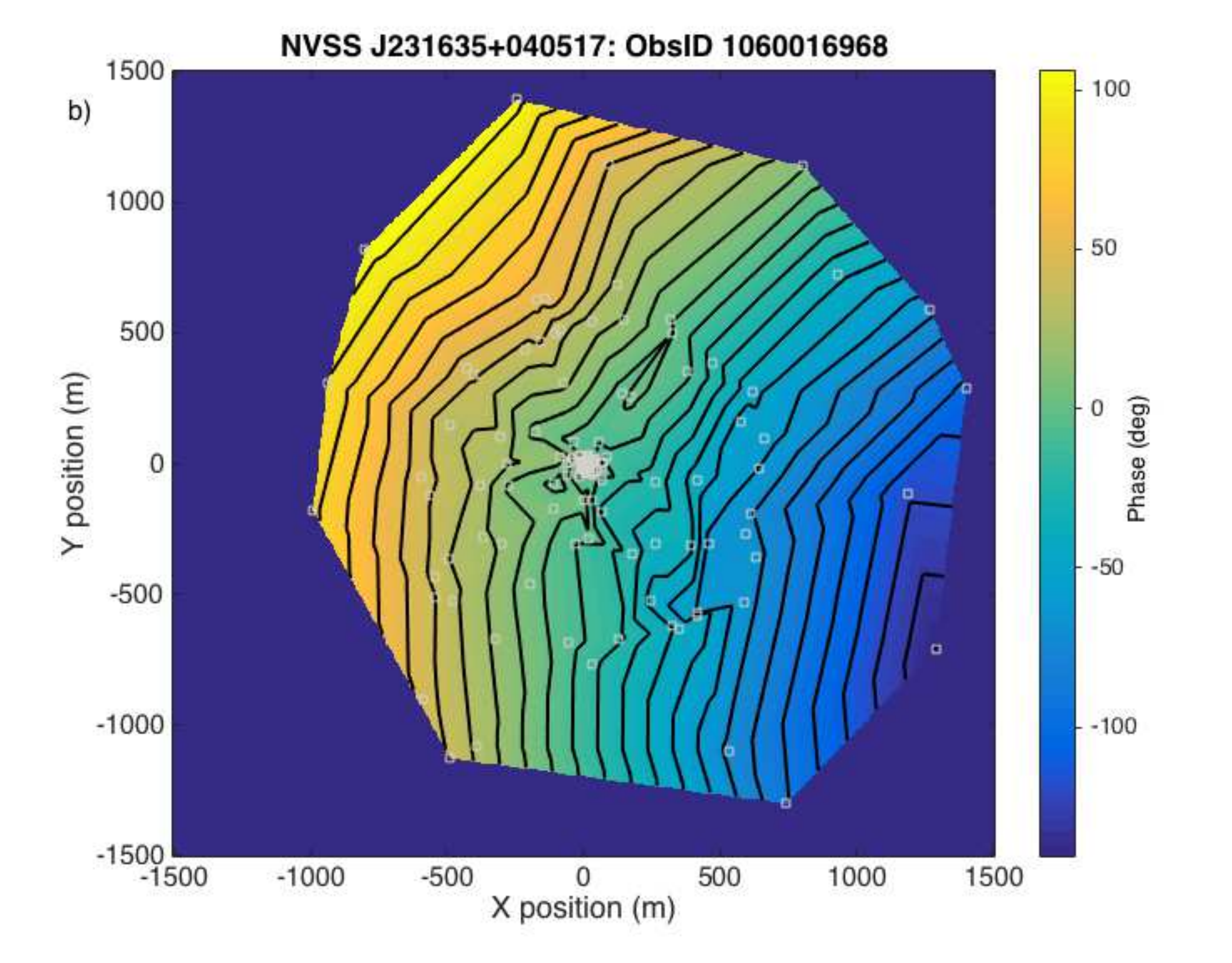}
\includegraphics[width=0.5\textwidth]{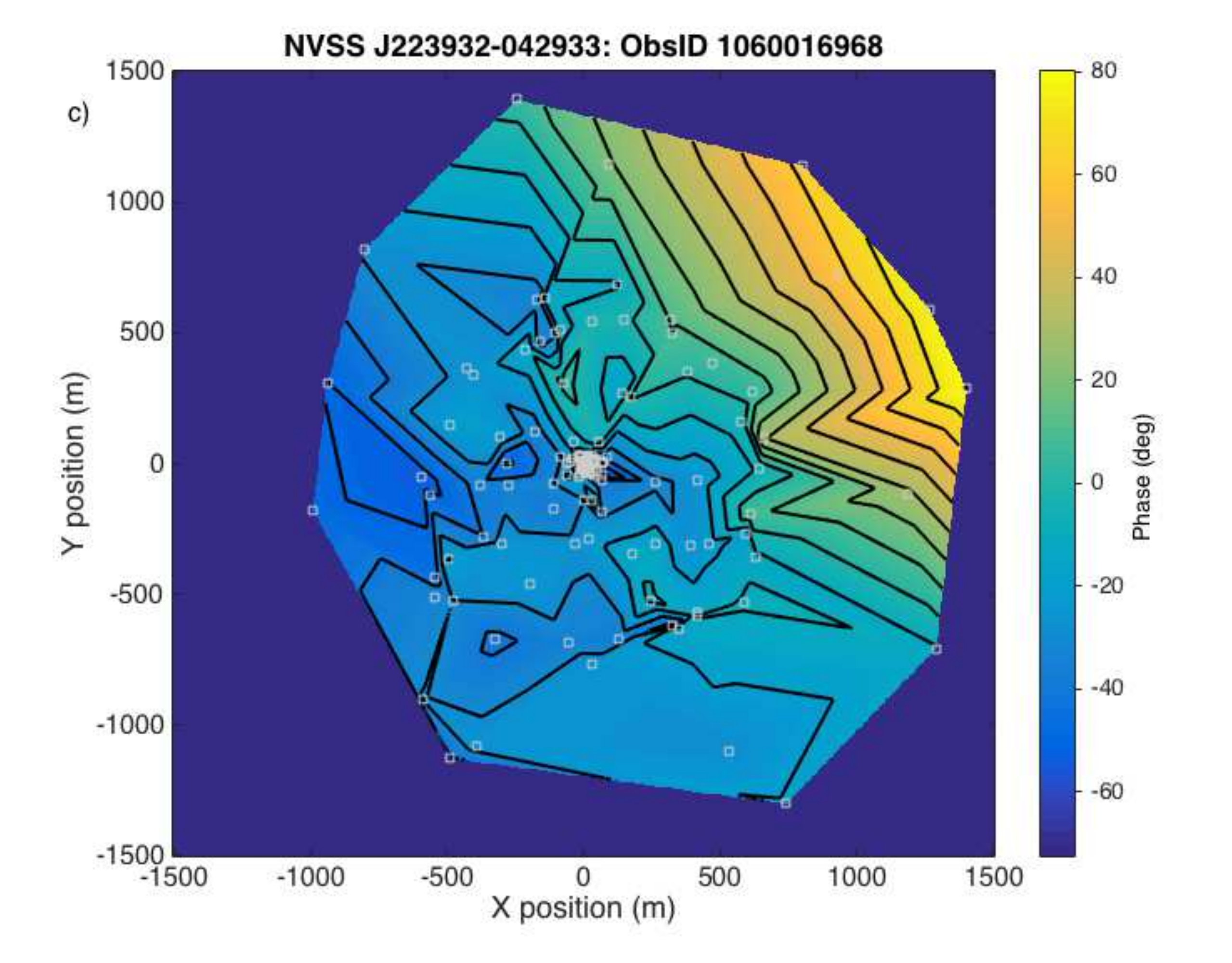}
\caption{Ionospheric phase screens over the MWA array, colour coded in degrees, with respect to the X-Y antenna locations, for a) moderate weather at 154MHz (Year-2 ObsID 1098800328), b) and c) for two different directions under poor weather conditions at 88MHz (Year-1 ObsID 1060016968). Contour spacing is 10\degr\ of phase and underlines the slowly changing approximately-planar phase screen over the array in a) and the steep and defocusing curved phase screens in b) and c). The ionospheric screen shown in b) is well described by a second order polynomial; instead for screen c)  the East-ward antennas show  a steep slope in the $\Delta$TEC above them, whilst those to the West show a flat phase screen, with a sharp transition between the two domains. 
MWA antenna positions are marked with grey boxes.\label{fig:ion_surf}}
\end{figure}

\begin{figure}
\centering
\includegraphics[width=0.6\textwidth]{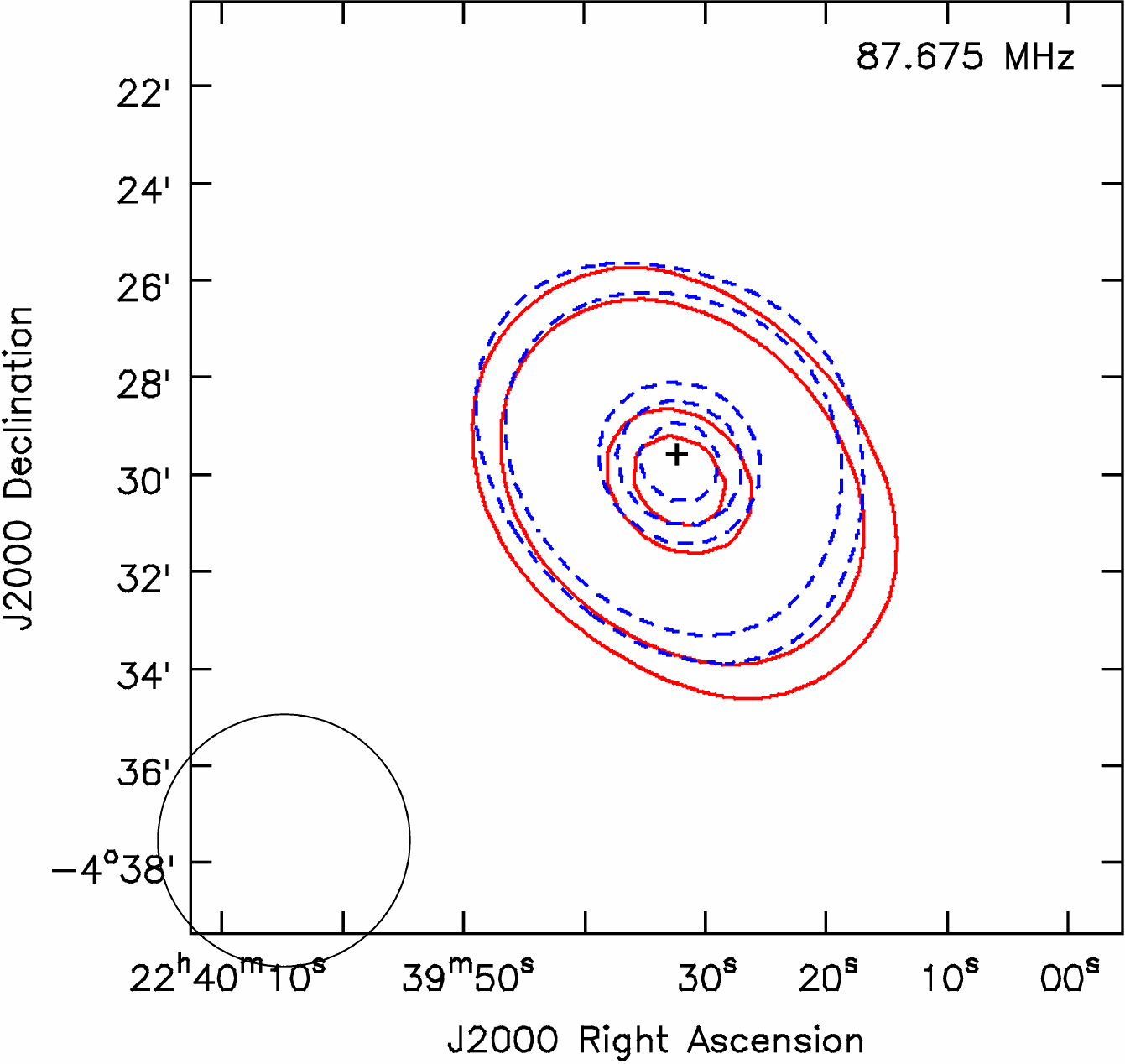}
\caption{
Example of ionospheric defocusing in a single snapshot image of \pfour\ under poor weather conditions (DI: red solid contour), at 88 MHz (Year-1, ObsID 1060016968). 
Overlaid is the LEAP single-direction full-calibrated image from the same observations (DD: blue dashed contour).
The contours are 2, 3, 8, 9 and 10\,Jy/beam. 
In the LEAP-calibrated image the peak intensity (10.7\,Jy/beam) is $\sim$10 per cent higher than in the original image (9.8\,Jy/beam), and the source is more compact (the plus sign marks the catalogue position for this source).
These DI-image artefacts result from the defocusing introduced by the curvature in the corresponding 
ionospheric phase screen over the array as measured with LEAP and shown in Fig. \ref{fig:ion_surf}c. These artefacts can not be corrected for by image-based ionospheric nor planar calibration techniques, as they can not reverse the source deformations.
\label{fig:ion_cons} }
\end{figure}

Here we summarise the typical and the extreme ionospheric conditions observed with the relatively short baselines of MWA, from the analysis in the directions of the LEAP calibrators within the FoV for the datasets considered in this paper.
For most cases the ionospheric screens can be approximated by a planar surface, which corresponds to the lowest order of ionospheric distortions. 
In this linear regime the sources in the images would be simply shifted, not deformed or defocused, and image-based corrections, as used in the GLEAM standard data analysis, would be equivalent to visibility-based LEAP calibration.
Note that the latter can also correct for shorter timescale variations.
Fig. \ref{fig:ion_surf}a shows the ionospheric screen measured in the direction of a strong (ca. 20 Jy) LEAP calibrator, which is precisely modelled by a planar surface.
However, results from weaker calibrators are less clear as the measurement noise is higher; 
it is easy to see that sensitivity is key to detect non-linear ionospheric distortions (i.e. higher order effects). 
Nevertheless significant curvatures in the ionospheric screens have been measured, even with the small footprint and limited sensitivity of MWA observations,
in periods of high ionospheric activity during GLEAM Year-1 observations, at 88\,MHz; because of this these were excluded from the standard analysis.
Examples of ionospheric screens are shown in Figures \ref{fig:ion_surf}b and c, in the direction of two strong LEAP calibrators.
The curvature in Fig. \ref{fig:ion_surf}b can be well described by a second order polynomial, whereas the surface in Fig. \ref{fig:ion_surf}c, for a different direction, does not match any simple functional form.
In these cases the image artefacts are, in addition to position shifts, source defocusing; therefore full calibration (i.e. non-planar) in the visibility domain is required.
Fig. \ref{fig:ion_cons} shows an example of the improvement from using LEAP full calibration with the phase screen shown in Figure \ref{fig:ion_surf}c with respect to image-based calibration, which provides $\sim$10 per cent increase in peak intensity due to the refocusing of ionospherically distorted radio signals in the single snapshot image.
This situation was unusual during the observations in GLEAM Year-1, with about 7 per cent of data affected. 
Nevertheless these effects will be more significant for observations with longer baselines, detectable at higher frequencies and under a wider range of weathers. Moreover, with the increased SKA-Low sensitivity such effects will be detectable even on short baselines.
Such issues, where they exist, would be corrected for with the LEAP calibration scheme.

The knowledge of the structure of the ionosphere  is vital for the characterisation of the power of the turbulent component in generic ionospheric models.
LEAP does not impose `a priori' constraints on the model of the ionosphere  (i.e. such as a functional dependence of the spatial structure) and therefore the measured antenna-based phase corrections are an unbiased probe of the ionospheric disturbances above the array, on a fine spatial scale.
These findings, at the site selected for SKA-Low, highlight the importance of DD calibration strategies that directly  sample the fine scale ionospheric disturbances, in multiple directions across the FoV.
LEAP provides a valuable resource for direct characterisation of the dominant source of propagation errors in the SKA era. 
This builds on previous work:
\citet{mevius_16} made a visibility-based analysis with the longer baselines of LOFAR, but restricted to directions of a few very strong sources; 
\citet{loi_15,jordon_17} used image-based techniques to probe variations in the spatial gradient over large  scales but are insensitive to ionospheric structures  smaller than the size of the MWA array.
The SPAM calibration method \citep{intema_phd,intema_09,intema_14}, like LEAP, produces a large dataset of ionospheric measurements based on antenna-based calibration solutions. SPAM furthermore projects the measurements to the height of the ionosphere and reconstructs the phase surface over the array, using the pierce points from different calibrators and antennas. However for MWA, unlike the VLA or GMRT, the pierce points do not overlap, so there is no benefit in doing a joint fit over multiple calibrators and we do not attempt to do so.


\subsection{Error analysis}\label{sec:error}

We have used simulation studies to estimate the errors introduced by calibrating with a  frequency-independent antenna phase value across 30.72\,MHz.
%
The residual phase errors  ($\phi$) after DI-calibration are expected to have a DD ionospheric-type component; i.e. $\phi_{\rm DD}(\nu) \propto -\Delta{\rm TEC}_{\rm DD}\,\nu^{-1}$.
Therefore, in principle, one could use the measured single-direction DDE set, along with the known 
wavelength dependence of the ionospheric phases, to apply the corresponding correction 
for every frequency channel across the whole bandwidth. In practise this was not possible. 
There is an additional large unknown geometric-type component ($\phi(\nu) \propto \nu$), introduced by the standard GLEAM Year-1 analysis, as detailed in Section \ref{sec:di}. 
The DI-calibration therefore incorporates the geometric-type flux-weighted average of the ionospheric source shifts across the FoV,  which is proportional to the flux-weighted average of the TEC gradients above the array. The corresponding phase contribution is proportional to this average gradient multiplied by the baseline length, which we indicate as $\overline{\rm TEC}$.
Consequentially the spectral signature of the inputs to the DD analysis,
which should have had a pure ionospheric-type signature, has an added unknown geometric-type component
(i.e. $\phi_{\rm DD}(\nu) \propto {\kappa \overline{\rm TEC}}\,\nu - \Delta{\rm TEC}_{\rm DD}\,\nu^{-1}$, where $\kappa \sim 1$) and we can not predict its functional dependence across the bandwidth.
This geometric-type contamination would not arise if the GLEAM Year-1 sky-model (or any other uncontaminated sky-model) was used for initial calibration, as is planned for the next release of GLEAM data (Franzen et al., in prep.). In that case, the geometric component would arise only from the catalogue position errors for the LEAP calibrators.
The typical coordinate errors in the MRC catalogue (3-10\,arcsecs), for example,  would not be significant.

In order to assess the effect of using the frequency-independent single-direction DDE set, we have carried out simulations using CASA, of MWA observations at 72--103 MHz with multiple realisations of phase errors, each with different flavours of spectral signatures arising from a steep phase gradient of two radians over the array. 
This gradient is the equivalent to a maximum $\Delta$TEC of 0.02\,TECU or $\sim$80\,arcsec position shifts, and was the limiting case in the FOM-1 studies. 
We generated synthetic datasets of MWA observations of a source with known flux and position, 
to which we added contaminating phase errors with either purely ionospheric, purely geometric or frequency independent phase signatures.
We proceeded with the analysis of the synthetic datasets as described previously in Section \ref{sec:dd}. 
In all cases we solved for the DDEs using the frequency-averaged datasets and apply them to the corresponding synthetic (unaveraged) dataset.
Finally we imaged the calibrated dataset, to measure the position shift and signal-loss of the synthetic source, for the three flavours of contamination. 

We find that the astrometric errors were under 2\,arcsec, the amplitude losses were less than 1 per cent, and no measurable increase in noise was observed in the calibrated image, compared to the uncontaminated image, in all cases.
Even better performance would be expected at higher frequencies where the ionospheric errors are smaller. Therefore we conclude that for observations with a Phase 1 MWA-type array  the errors resulting from neglecting the frequency dependence of the DD phase calibration, in our current implementation of LEAP, are small in all cases.
The breakdown of the use of frequency-independent antenna phase values will commence when there are significant phase variations across the band;       
those will arise in observations under larger $\Delta$TEC values or, equally, longer baselines.
While this limit would only be reached with MWA under unusually bad weather conditions, with the longer baselines of MWA Phase 2 it is expected that this will happen more often. 
LOFAR has longer baselines and wider bandwidths, so will reach this limit under most weather conditions.

In general, the ionospheric effects manifest in different image artefacts depending on the sizes of both the array and the FoV relative to the spatial scale of phase structure in the ionosphere. This leads to four calibration regimes of wide or narrow FoV, on short or long baselines \citep{lonsdale_05,intema_09}. 
A natural extension of our method that addresses the issues above and allows operation in all calibration regimes would involve replacing the uniform averaging with baseline dependent averaging. 
We have taken s.pdf to upgrade our current strategy to introduce these developments. 
This will result in a more effective directional filter, whilst preserving the spectral signature of the antenna-based phase corrections; 
this is important to reduce the propagation of errors when applying the corrections to long-baseline datasets. 
The updated strategy will retain the embarrassingly parallel processing nature and the requirement for a limited local sky model, 
that is vital for efficient DD calibration.

\subsection{Availability of calibrators} 
\label{sec:cals}

\begin{figure}
    \centering
    \includegraphics[width=0.7\textwidth]{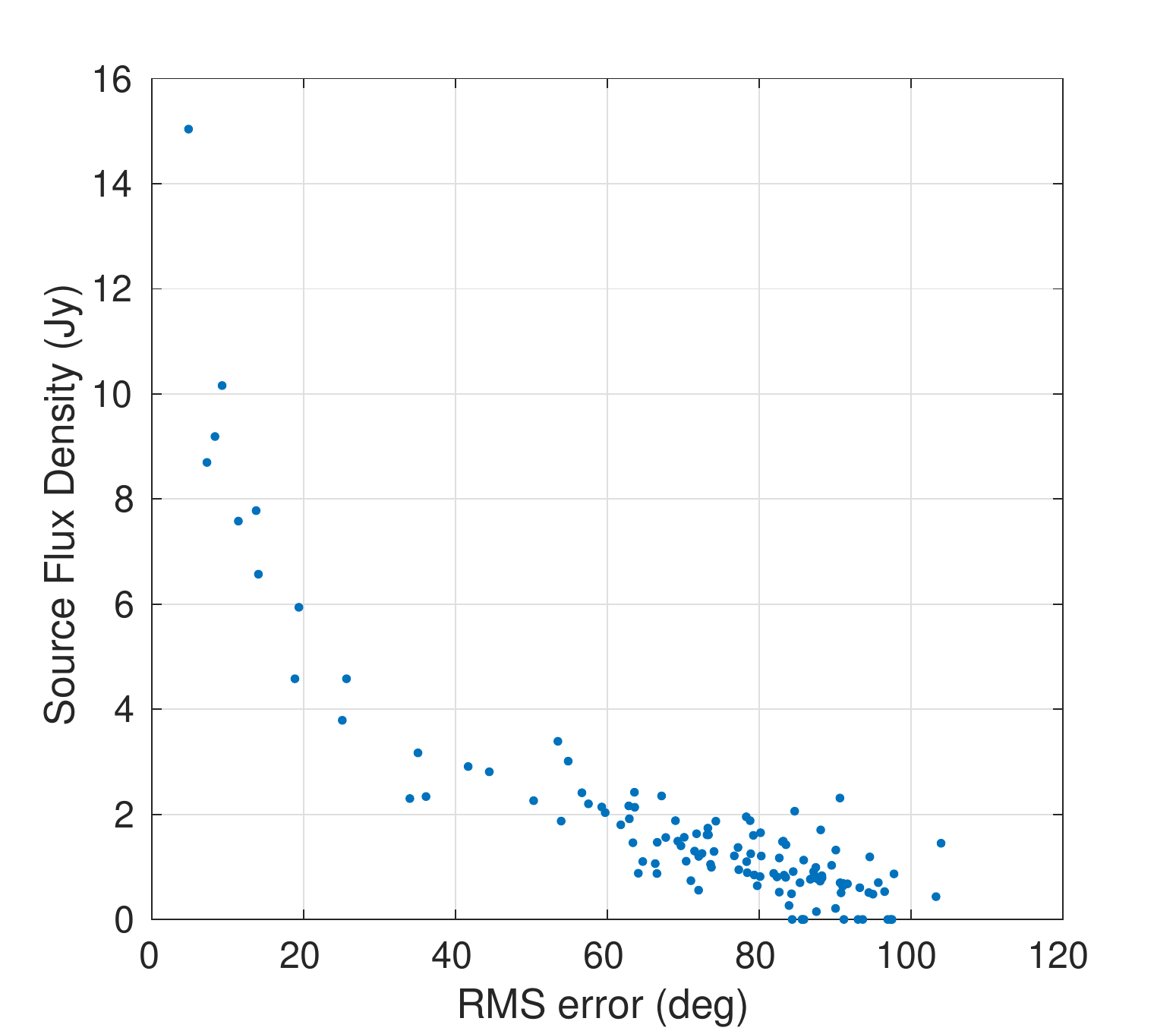}
    \caption{Apparent flux densities for all MRC sources in the verification field (15\degr$\times$15\degr), against the corresponding rms error of the planar fit to the DDE surfaces for each source, for a single snapshot at 154MHz. 
    The transition between the domains where the rms error is dominated by the calibrator source flux density, $S_{\rm cal}$, to that where the rms error is dominated by the flux density of the smeared background sources, $S_{\rm b}$, falls approximately at 2\,Jy or 60\degr\ of rms error. \label{fig:flx_rms}}
    \label{fig:flux_surfacerms}
\end{figure}

\begin{figure}
    \centering
    \includegraphics[width=0.7\textwidth]{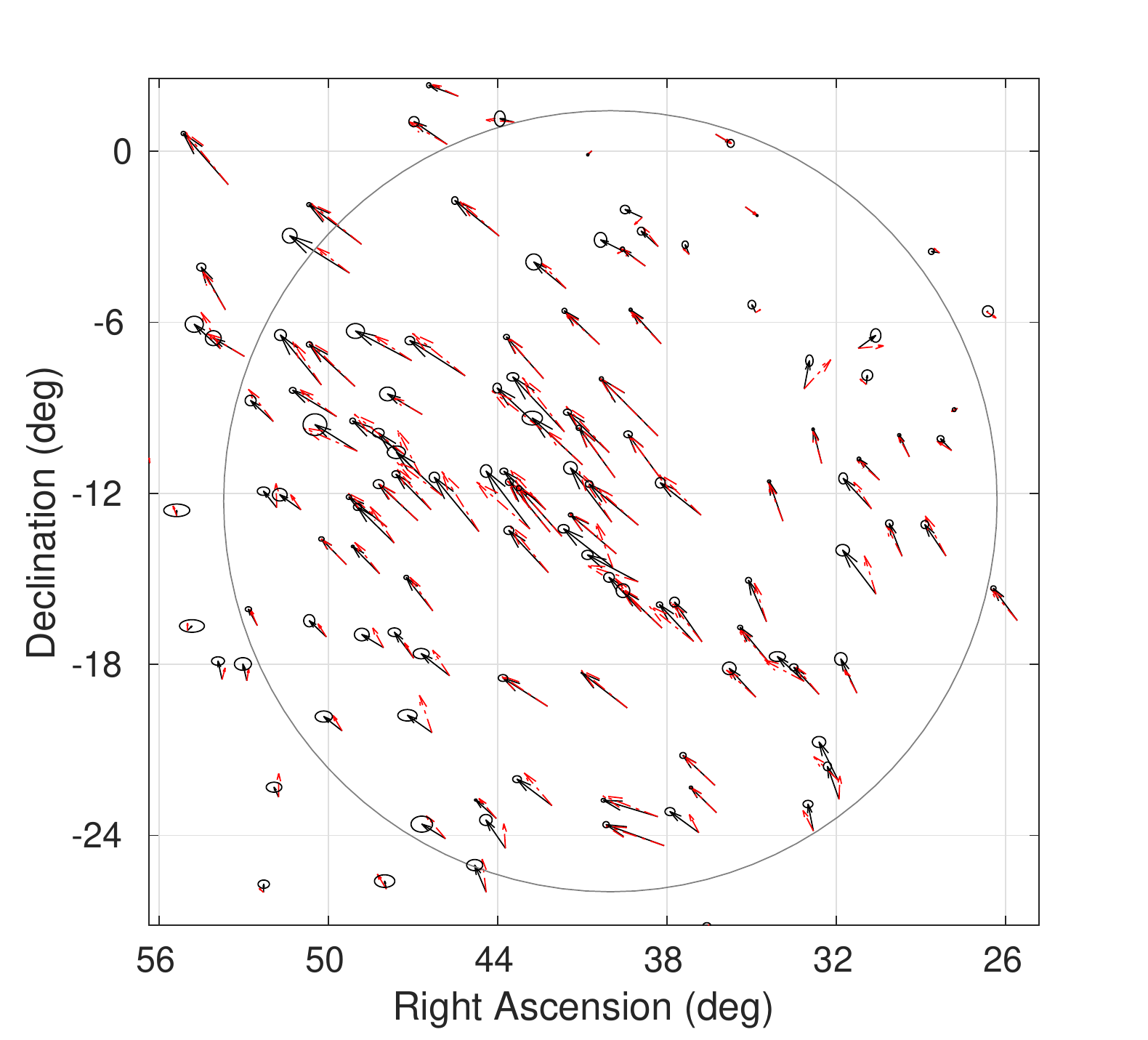}
    \caption{Spatial distribution of 70 identified LEAP calibrators among the 450 MRC sources within the full FoV of the last snapshot in the verification dataset, at 154\,MHz. 
    The LEAP calibrators have a rms error in the planar fit to the DDE surfaces <60\degr. The position shifts inferred from the LEAP calibration in the visibility domain are shown as dotted vectors in red. The  position shifts measured in the image domain are shown as vectors in black. The 1$\sigma$ errors in the positions are shown as a small circle. The MWA primary beam response is marked in grey, outside of which few sources are detected.}
    \label{fig:sol_vec}
\end{figure}

The general requirements of a LEAP calibrator are described in Section \ref{subsec:leapcals}.
Here we present three additional approaches, complementary to the FOM studies, to determine the availability of LEAP calibrations. 
The first two are empirical approaches based on the results from the LEAP analysis, in the direction of all the MRC sources within the FoV, for all the MWA observations considered in this paper (Section \ref{sec:obs}). They assess the performance of the LEAP directional filter and the SNR threshold.
Furthermore we predict the LEAP calibrator sky-coverage, using the GLEAM catalogue.

The first approach consists of using the rms errors of the planar fit to the DDE surfaces, as a measure of the phase noise contribution from the smeared sources in other directions. 
In general the rms errors will have two contributions: from the signal in the direction of interest, and from the smeared sources in other directions. 
The former will dominate for strong calibrators, and is expected to follow an inversely proportional relationship,  as rms $\propto S_b/S_{\rm cal}$, with $S_b$ the flux density from smeared sources and $S_{\rm cal}$ the calibrator flux density \citep{kassim_07}.
Fig. \ref{fig:flux_surfacerms} shows the apparent flux densities measured in a MWA snapshot image, versus the rms error, for the 126 MRC sources within the {15\degr$\times$15\degr} verification field. 
Deviations from the inversely proportional trend, as seen for weaker calibrator sources in Fig. \ref{fig:flux_surfacerms}, are indicative of the significant contribution from other directions, breaking the condition to be a LEAP calibrator.
The plot shows a change in trend for apparent flux densities of ca. 2\,Jy, which corresponds to an rms error of 60\degr. This marks the transition between the two regimes, calibrator dominated versus smeared sources dominated, and sets a lower flux density threshold for the LEAP calibrators. 
In this field there are 25 LEAP calibrators, out of the 126 potential sources, which have planar fit rms errors was $<$60\degr\ and have apparent flux densities $>$2 Jy. 
Typically we find around 100 LEAP calibrators among the 300 to 800 MRC sources in the MWA FoV.

The second approach to identify LEAP calibrators, which supports the first,
is to compare the DD residual errors measured in the visibility and image domains.
Both should agree for LEAP calibrators in the calibration regime of MWA (i.e. short baselines, wide FoV).

In order to do this, we used the planar fit to the DDE surfaces to calculate the implied effects in the image domain, using the conversion between phase gradients in the X-Y plane and position shifts in the image domain.
Those are compared to the offsets measured in the image domain, between the source positions in the DI-calibrated snapshots with respect to positions in the MRC catalogue. 
This was done for all the MRC sources within the FoV, for all snapshots.
In general, good alignment between both vectors was found when the planar fit rms error $<$60\degr. This agrees with the selection criteria from the first approach, being equivalent to a lower flux density threshold of 2\,Jy, as shown in Fig. \ref{fig:flx_rms}.
Fig. \ref{fig:sol_vec} shows an example of the distribution of LEAP calibrators (selected as described above) within a snapshot FoV and the corresponding position shift vectors measured in both domains. 
The grid superimposed in the figure has a cell size of 6\degr$\times$6\degr; there are 70 LEAP calibrators, out of the 450 MRC sources, 
and no grid square within the FoV is without a calibrator. 
Hence, the calibrator density surpasses the requirements from the FOM analysis, which sets a maximum calibrator separation of 6--10\degr. 
These findings are representative of those for other datasets and regions of the sky.
The exception to this general trend is when an extremely strong source lies in the FoV, which overwhelms other potential calibrators in a region around it. 
For example one of the Year-1 datasets in our analysis included Fornax-A, which has an apparent flux density of ca. 400\,Jy at 88 MHz.
We find that such a source creates a void around itself, suppressing any other LEAP calibrator, out to a radius of ca. 10\degr. 
However, even in this extreme case 60 per cent of the sky in this void is within 3\degr\ of a LEAP calibrator and 100 per cent is within 5\degr\ of a LEAP calibrator.
Alternatively, to handle such cases, one could opt for ``peeling'' the limited number of the extremely bright sources when they lie in the FoV. 

The third approach is to make a theoretical estimate of the availability of LEAP calibrators across the whole observed sky at both frequencies. We have used the source positions and flux densities
listed in the GLEAM catalogue and calculated, for each source, the two SNRs relevant for a LEAP calibrator (Section \ref{subsec:leapcals}). These are: 
1) the ratio of its flux density over the typical MWA snapshot image noise, and 
2) the ratio of its flux density over that from the combination of the smeared sources around it. 
For the latter, we have calculated the smeared flux densities, resulting from averaging a bandwidth of 30 MHz in observations with a MWA-scale array, for all other sources within a region 40\degr,
and combined them in quadrature. We find that two thirds of the sky has a LEAP calibrator within 3\degr, if we require both SNRs to be greater than five; for SNRs greater than three, more than 80 per cent of the sky has a LEAP calibrator within 3\degr. 
The excluded zones arise from a few of the very strongest sources. Subtracting these sources from the visibilities would allow for full sky coverage. 

\section{Conclusions}\label{sec:conc}
\subsection{The LEAP calibration scheme}

We have presented an end-to-end demonstration 
of the DD calibration scheme LEAP, using MWA observations at 88 and 154 MHz, under a range of weather conditions. 
%
The DD errors are measured and corrected for in the visibility domain, which is the optimal place for antenna-based corrections. 
The images are generated using existing wide FoV imagers, so little new software is required. 

Since each direction is independent, LEAP is unique in its embarrassingly parallel computing performance, as it does not require a global sky model. In contrast, nearly all other DD-calibration approaches are sequential or simultaneous processes, neither of which are ideal for the SKA Science Data Processing. 

LEAP can mitigate the effect of linear and fine scale higher order ionospheric phase distortions, smaller than the size of the array, as well as DD antenna-beam response differences.
It allows for more robust calibration solutions and therefore a higher fidelity imaging under a wider range of weather conditions, compared to image-based corrections. 

We have presented performance evaluation studies to assess the effectiveness of the directional filter, the validity of the calibration solutions, the calibrator availability and the errors introduced. 
These show that the accuracy of the calibration suffers from an approximately linear degradation 
away from the direction of the LEAP calibrator, reaching the level of the DI calibration at directions ca. 3--5\degr\ away,  at both 88 and 154 MHz. 
%
This determines the maximum size of the facets to be used during imaging to achieve a net-improvement, which we used to produce a DD-mosaiced image. Further increases in quality will come from using smaller facets, either from increasing the number of calibrators or by using spatial interpolation for intermediate directions.
Also, our verification tests indicate that the availability of LEAP calibrators 
is not an issue over the majority of the sky, even with the limited sensitivity of MWA.
We have demonstrated the feasibility of LEAP for MWA-scale arrays, in the regime of short baselines and wide FoVs.

\subsection{Extrapolation to long baselines and future instruments} \label{sec:conc_long}

LEAP is equipped for the challenges posed by longer baselines, i.e. higher order distortions and uncorrelated ionospheric effects over the array. 
While our demonstration is in the short-baseline calibration regime, which is dominated by first order effects that can be corrected in the image domain, even here there are hints of higher order spatial ionospheric distortions and short timescale variations;
such effects are expected to be more significant with increased sensitivity. 
We expect that the LEAP strategy presented in this paper will continue to be useful with longer baselines and under a wide range of weather conditions, for LOFAR, MWA Phase-2/3 and SKA-Low, when combined with baseline dependent averaging. 

\subsection{Fine scale ionospheric structure} 

The knowledge of the structure of the ionosphere is vital for the characterisation of 
ionospheric models. 
As LEAP does not impose `a priori' constraints 
the measured antenna-based phase corrections are an unbiased probe of the ionospheric disturbances above the array, on fine spatial and temporal scales, for all directions across the FoV. 
Therefore LEAP characterises the fine scale ionospheric phase structure, which is the most intractable and poses a major challenge for future instruments and their science goals.

\section*{Acknowledgements}
We acknowledge with thanks the report from the anonymous referee, which significantly improved this paper.
This scientific work makes use of the Murchison Radio-astronomy Observatory, operated by CSIRO. We acknowledge the Wajarri Yamatji people as the traditional owners of the Observatory site. Support for the operation of the MWA is provided by the Australian Government (NCRIS), under a contract to Curtin University administered by Astronomy Australia Limited. We acknowledge the Pawsey Supercomputing Centre which is supported by the Western Australian and Australian Governments.



\bsp	
\label{lastpage}
\end{document}